\documentclass[preprint,12pt]{elsarticle}
\usepackage{amssymb}
\usepackage{booktabs}
\newcommand{\fedsim}{\textit{FedSim}\xspace}
\newcommand{\fedavg}{\textit{FedAvg}\xspace}
\newcommand{\fedprox}{\textit{FedProx}\xspace}

\newcommand{\greads}{Fed-Goodreads\xspace}
\newcommand{\mex}{Fed-MEx\xspace}
\newcommand{\feddct}{\textit{FedFT}\xspace}
\newcommand{\fspace}{the frequency space\xspace}

\usepackage{xspace}
\usepackage{algorithm}
\usepackage{algorithmic}
\usepackage{subcaption}
\usepackage{amsmath}
\usepackage{cancel}
\usepackage{multirow}
\usepackage{xcolor}         % colors

\begin{document}

\begin{frontmatter}

\title{FedFT: Improving Communication Performance for Federated Learning with Frequency Space Transformation}

\author[label1]{Chamath Palihawadana\corref{cor1}}
\ead{c.palihawadana@rgu.ac.uk}
\cortext[cor1]{Corresponding author.}
\author[label1]{Nirmalie Wiratunga}
\author[label2]{Anjana Wijekoon}
\author[label1]{Harsha Kalutarage}
\affiliation[label1]{organization={School of Computing, Engineering, and Technology, Robert Gordon University},
            addressline={Garthdee Rd}, 
            city={Aberdeen},
            postcode={AB10 7AQ}, 
            state={Scotland},
            country={United Kingdom}}
\affiliation[label2]{organization={University College London},
            addressline={Gower St}, 
            city={London},
            postcode={WC1E 6BT}, 
            state={England},
            country={United Kingdom}}

\begin{abstract}
    Communication efficiency is a widely recognised research problem in Federated Learning (FL), with recent work focused on developing techniques for efficient compression, distribution and aggregation of model parameters between clients and the server. 
    Particularly within distributed systems, it is important to balance the need for computational cost and communication efficiency.
    However, existing methods are often constrained to specific applications and are less generalisable. In this paper, we introduce \textit{FedFT} (federated frequency-space transformation), a simple yet effective methodology for communicating model parameters in a FL setting. \textit{FedFT} uses Discrete Cosine Transform (DCT) to represent model parameters in frequency space, enabling efficient compression and reducing communication overhead. \textit{FedFT} is compatible with various existing FL methodologies and neural architectures, and its linear property eliminates the need for multiple transformations during federated aggregation. 
    This methodology is vital for distributed solutions, tackling essential challenges like data privacy, interoperability, and energy efficiency inherent to these environments.
    We demonstrate the generalisability of the \textit{FedFT} methodology on four datasets using comparative studies with three state-of-the-art FL baselines (\textit{FedAvg}, \textit{FedProx}, \textit{FedSim}).
    Our results demonstrate that using \textit{FedFT} to represent the differences in model parameters between communication rounds in frequency space results in a more compact representation compared to representing the entire model in frequency space.
    This leads to a reduction in communication overhead, while keeping accuracy levels comparable and in some cases even improving it.
    Our results suggest that this reduction can range from 5\% to 30\% per client, depending on dataset.   
\end{abstract}

% \begin{highlights}
% \item Research highlight 1
% \item Research highlight 2
% \end{highlights}


\end{frontmatter}

%% \linenumbers

\section{Introduction}
\label{sec:intro}

Federated Learning~(FL) plays a crucial role in advancing decentralised Artificial Intelligence~(AI) training, 
as it allows for the training of Machine Learning (ML) models on distributed clients (i.e. edge devices) without centralising sensitive data.
Training of ML models on client data without transferring them to a central server significantly reduces the risk of privacy violation and ensures better compliance with regulations such as the GDPR~\cite{albrecht2016gdpr}.
FL seamlessly integrates with distributed systems, emphasising distributed processing and enhanced data privacy. This method trains machine learning models directly on edge devices, cutting down bandwidth needs and significantly improving privacy measures \cite{bao2022federated}.
The prevalence of FL applications, such as voice assistants, Internet of Things (IoT) devices and mobile apps are %also 
rapidly increasing
~\cite{wakeword,emoji,gboard,nguyen2021federated}.
Other areas such as privacy-sensitive ML applications in healthcare for disease diagnosis and treatment are also ideal domains for FL~\cite{chen2020fedhealth,healthcarepredict, healthrecords}. 
Another notable industry adapting FL is in finance, where models can be trained without shared access to private information for credit risk assessment and anti-fraud detection~\cite{long2020federated}.

\begin{figure}[htb]
\centering
\includegraphics[width=1\textwidth]{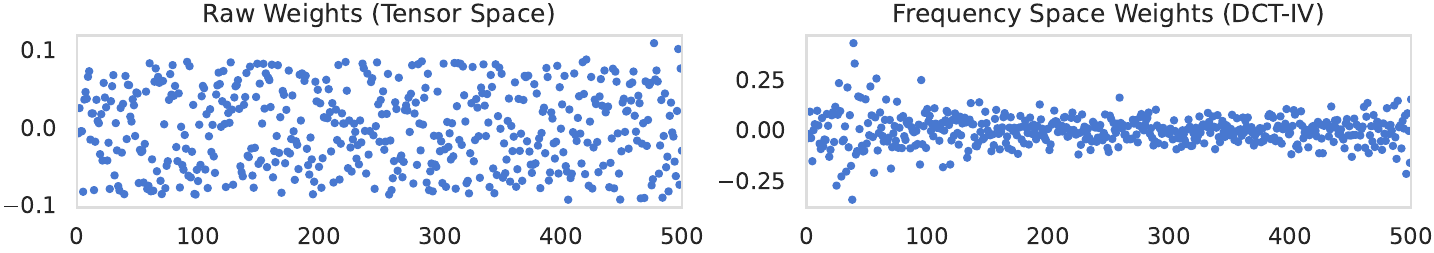}
\caption{Model parameters represented in tensor space and frequency space.}
\label{fig:orignalvsfreq}
\end{figure}

Central to FL is its distributed decentralised training of a shared global model with many communication exchanges between the server and its clients. 
At each communication round, the shared model is updated by the server as an aggregation of client models received. 
The FL communication layer needs to handle many requirements due to its iterative nature which involves frequent and large model exchanges.
Inefficient communication can slow down training, increase computational cost, decrease accuracy, raise energy consumption and limit scalability~\cite{konevcny2016federated}.
Compression can be used to improve communication performance by reducing the amount of data being transmitted. 
For instance, ML applications can optimise storage and inference speed by using transformation methods like Discrete Cosine Transform (DCT), as demonstrated in~\cite{liu2018frequency}. 
DCT operates by converting model parameters into the frequency domain, after which processes like quantisation and pruning can be applied to discard less significant coefficients. 
This results in a more compact representation of the model with optimising both storage requirements and computational efficiency during inference.
In FL research, although the use of DCT has been acknowledged, 
the main focus has been on using it to compress training data for better local representation on client devices~\cite{han2021application,chen2022fltalk}. 
We use the term ``tensor space'' to distinguish the space in which model parameters are represented from that in which training data is represented. 
The example in Figure \ref{fig:orignalvsfreq} compares model weight representation in tensor and frequency spaces. In the frequency space, the weights decomposed into their constituent frequencies are more spread out and concentrated to a few dominant frequency components, allowing for efficient representation and storage.  

One of the challenges facing FL using tensor space compression is ensuring that compression techniques do not obstruct server-side aggregation operations.
Most FL methods address this challenge by using lossless compression techniques and incorporate an extra step of reconstructing the tensor space at the server for model aggregation prior to compressing it again for transmission back to the clients~\cite{felix2020,dai2019hyper}.
What we propose in this paper is a methodology, \feddct, that enables server aggregation in the same compressed space.
To achieve this, we investigate the feasibility of using DCT-transformed model parameters in the communication layer of FL to enhance communication performance without sacrificing model accuracy. 
The direct advantage of \feddct is that it enables the sharing of model parameters in the frequency domain, and local client updates can be done in either \fspace or tensor space, making it adaptable across different methodologies.
Furthermore, the compact representations in frequency space make it simple for clients to identify sparse areas that could be easily pruned before communicating them to the central server.

\noindent Accordingly, we make three contributions.
\begin{itemize}
\item Introducing \feddct, a novel FL methodology that utilises frequency space transformation to improve communication efficiency while preserving performance.
\item Conducting a comparative study with state-of-the-art FL methodologies to demonstrate the generalisability of the frequency space transformation and evaluate the trade-off between model performance and communication efficiency.
\item Demonstrating the generalisability of \feddct by analysing evaluation results from a range of neural architectures for image, text and sensor data.
\end{itemize}

The rest of this paper is organised as follows. 
Related work is discussed in Section~\ref{sec:background} followed by the proposed \feddct methodology presented in Section~\ref{sec:methods}. 
The role of model variance in transformed communication is discussed in Section~\ref{sec:fspace}. 
Experiment setup and evaluation scenarios appears in Section~\ref{sec:eval} and results are discussed in Section~\ref{sec:results}. 
Conclusions and future directions are presented in Section~\ref{sec:conc}.

\section{Background and Related Work}
\label{sec:background}

The exponential growth of edge devices and their applications, including the IoT, smart home assistants, mobile apps, and wearables, has led to a substantial increase in data creation at the edge of the network. 
This increase of data (i.e. big data) offers considerable benefits for customisation and real-time analytics but also introduces significant risks to user privacy.
Accordingly, the demand for distributed machine learning strategies, FL and distributed computing architectures has intensified, becoming more crucial than ever.
In practical applications FL operates by training models on local data directly on client/edge devices, emphasising privacy and minimal data transfer \cite{imteaj2021survey}. 
In contrast, distributed architectures focus on processing data locally at the edge of the network first and if required communicate it to a central cloud for further processing. 
The key similarity between FL and distributed computing lies in their reliance on local computation to reduce bandwidth usage. 
Both strategies are designed to efficiently manage data where it is generated, thereby reducing network resources and enhancing communication efficiency \cite{bao2022federated}.

\subsection{Communication Efficient FL}
Communication cost is a primary bottleneck for FL systems~\cite{zhao2023towards,survey1,survey2}.
This is because FL requires high-frequent communication of model parameters between clients and the server, where the number of clients can be in the millions~\cite{niu2020billion,bonawitz2019towards}. 
The size of these neural models can vary greatly, from a mere few kilobytes to several hundred megabytes depending on their complexity.
The communication bottleneck in FL systems can lead to unreliability and limit their ability to scale up and meet increasing demands.
In a realistic setting, there can be clients with poor network connections or with resource limitations which can further hinder the performance of the FL system. 
Methodologies like \fedprox~\cite{fedprox} have addressed this issue to handle partial updates from clients having limited connectivity.
More generally, the approaches in literature aimed at mitigating this communication bottleneck can be studied under two groups: \textit{structured updates} - where the local training and communication is done in a restricted space~(e.g. restricted to a fewer number of model parameters); and \textit{sketched updates} - where the local update is performed on the complete model and compressed for communication~\cite{konevcny2016federated}. 
Both approaches have advantages and disadvantages, however, the efficiency of communication in updates performed through sketches is often more adaptable and can be easily integrated with existing FL techniques.
We employ the sketched update approach to enable local client updates to be performed in either tensor or compressed frequency space, with communication and federated aggregation performed in compressed frequency space. 

\subsection{Compression for Communication Efficiency}
\begin{figure}[htb]
\centering
\includegraphics[width=\textwidth]{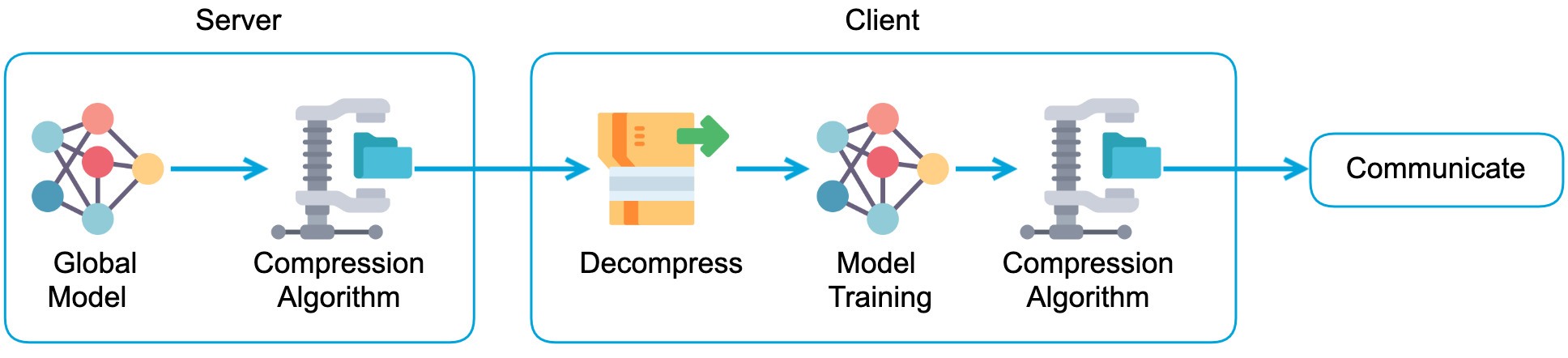}
    \caption{Process of applying standard compression algorithms}
    \label{fig:compressionssample}
\end{figure}

Compression techniques of FL models in tensor space include sub sampling (which reduces spacial resolution) and probabilistic quantisation~(which reduces precision)~\cite{konevcny2016federated}. 
The aim of our work is closer to~\cite{felix2020}, where they compress model parameters using Golomb Coding and reduce model complexity through quantisation. 
Golomb Coding is a lossless compression technique, but its non-linear nature means that additional transformations must be performed at the server incurring extra reconstruction steps. This is because it cannot perform federated aggregation in the compressed space. Also, the quantisation is tightly coupled with the client local update making the work by~\cite{felix2020} less adaptable by existing FL methodologies. 
Figure \ref{fig:compressionssample} illustrates applying a standard compression algorithm in FL.
This approach involves multiple stages of model compression and decompression to reduce the communication overhead between the server and the clients.
The increased computational demand can make the approach less practical, especially for resource-constrained environments or large-scale deployments with millions of clients. 

The proposed \feddct alleviates both shortcomings by choosing a linear and orthogonal transformation technique like DCT (Type $IV$ ~\cite{dctiv}) where federated aggregation can be performed in \fspace and the methodology is decoupled from general FL. 
Specifically its orthogonality property is essential for compression and its linearity property enables federated aggregation on \fspace.
Authors of~\cite{dai2019hyper}, also aim for communication efficiency 
through lossless compression and quantisation, but they apply compression on the gradients rather than model parameters. 
\cite{dai2019hyper} share the same limitation of adaptability for FL as discussed for~\cite{felix2020}. 

Frequency space transformation techniques have been employed for data compression for many years, examples include DCT, Discrete Fourier Transform (DFT), Fast Fourier Transform (FFT), and Principal Component Analysis (PCA).
However, DCT is one of the most widely used techniques due to its favourable properties such as computational efficiency and the ability to compactly represent the energy content of a signal~\cite{rao2014discrete, strang1999discrete}. 
As a result DCT is widely adopted for image and video compression applications. 
Data compression in the frequency space is accomplished by pruning (i.e. trimming) or quantising the least significant information.
In the field of ML, researchers have explored using the DCT for compressing models by transforming data into the frequency space~\cite{robinson2003combining, dimililer2022dct}. 
To the best of our knowledge, the full potential of DCT has not yet been fully exploited in the context of FL for the purpose of improving communication efficiency through model compression.

\subsection{Pruning and Quantisation for Communication Efficiency}

Pruning and quantisation are two methods used to optimise and simplify ML models~\cite{liu2018frequency,jiang2022model,zhao2023towards}.
Pruning removes elements that carry less significant model parameters to reduce model size; a pruning mask can be learnt as part of model optimisation~\cite{liu2018frequency}; alternatively,  prior knowledge is used to identify pruning areas in static pruning. 
Quantisation aims to decrease the precision of model parameters through the use of fewer bits, resulting in a smaller model size. Both techniques play a crucial role in ML models deployed in resource-constrained environments. Some quantisation techniques used in FL include FedPAQ~\cite{fedpaq} using periodic averaging and FedPara~\cite{fedpara} using low-rank Hadamard product parameterisation to reduce the precision of the model weights. 
Authors of~\cite{prakash2022iot} adopt a similar method to~\cite{liu2018frequency} in FL to learn a pruning mask during client training which reduces the upstream communication cost. In contrast, a global and client model pruning is applied by~\cite{jiang2022model} using static pruning masks to reduce the overall communication cost. Their approach needs an extra step at the server where a single gradient descent step is taken on the global model.
\feddct utilises prior knowledge of pruning in \fspace, eliminating the need for learning pruning masks and providing a straightforward method to reduce communication costs by incorporating ideas from compression but in the frequency space.

\subsection{Aggregation Techniques in Federated Learning}

The fundamental principle of FL revolves around the concept of federated aggregation. This process involves merging local model updates from distributed clients to construct a unified global model, a step of critical importance in diverse settings such as distributed architectures.
Federated Averaging (\fedavg) is the conventional FL algorithm which was introduced by \cite{fedavg}. The aim of \fedavg is to create a effective global model with wider coverage from the participating clients. 
This is achieved through a weighted aggregation approach, where the influence of each client on the final model is proportionate to the size of its data. In practical terms, this means that clients with larger datasets have a greater impact on the FL system, a crucial consideration in distributed environments where data volume can vary significantly among clients.
\fedavg effectively harnesses the collective data wealth of all participating nodes, leading to a more holistic and representative global model in diverse FL scenarios.

Another state-of-the-art FL methodology is the \fedprox algorithm, introduced by \cite{fedprox}, offers practical solutions for handling the challenges of system and statistical heterogeneity in FL.
These challenges are particularly evident in distributed computing and distributed machine learning scenarios.
\fedprox is effective in environments where clients vary in computational power and network connectivity. It allows for partial updates from clients, a feature that is especially useful in distributed systems where devices might not always complete their computations due to varying capabilities or connectivity issues. The algorithm measures the extent of computation each client manages to complete, making it adaptable to the inconsistent participation often seen in these systems.
For statistical heterogeneity, \fedprox introduces a proximal term in the local model update. This term acts like a balancing factor, ensuring that local models do not stray too far from the global model, despite the diverse data they might be learning from. This is particularly useful in distributed setups where each client might have access to very different types of data.

While \fedavg emphasises weighting clients according to their data distribution and \fedprox effectively manages partial updates, the \fedsim \citep{fedsim} algorithm introduces a distinct strategy: a similarity-guided clustering approach. During each iteration, \fedsim clusters clients based on the similarity of their local data, which is inferred from the shared, updated model. This initial phase involves performing an aggregation at the cluster level, primarily based on the size of the data. 
Where each cluster synthesises a local cluster model, and these models are then collectively averaged to build the global model. 
This method allows for a more detailed and targeted approach, which is particularly useful in networks where clients with similar data are spread out. It lets clients with similar data work together closely, while also making sure that clients with different data can contribute in their own way. This balanced way of working improves both the efficiency and effectiveness of the learning process in networks where data is shared across many clients.
This method is particularly effective in distributed systems, where clients often have similar data characteristics. 

All the aggregation techniques discussed possess unique advantages in the context of FL, yet they converge on a critical aspect: the communication bottleneck. This bottleneck arises from the necessity for each client to transmit updated models back to the central server \cite{caldas2018expanding, chen2021communication, pouriyeh2022secure}. 
Addressing this crucial need, our focus shifts to enhancing communication within FL. This is where our proposed \feddct method comes into play, aiming to significantly improve communication efficiency in these environments.

\section{\feddct Methodology}
\label{sec:methods}

FL is a distributed machine learning approach where models are trained locally on individual devices or nodes, and only aggregated model updates are shared, preserving data privacy and minimising communication costs. It enables collaborative learning without centralised data collection.
The general FL process begins at round, $t=0$, with a server distributing an initial global model, $w_0$, to all participating clients.
At each communication round $t$, the server selects $K$ clients to participate in training. 
Clients perform local training, and once complete, each client, $k$, communicates their model $w_{t+1}^k$ to the server.
These models are aggregated as in Equation~\ref{eq:fedavg} to form the global model at $t+1$. This is repeated for multiple communication rounds. 
The Federated Averaging algorithm~\fedavg~\cite{fedavg}, computes a weighted average of locally trained client models, where the weights are determined by the size of each client's data $n_k$.
\begin{equation}
 \begin{split}
    w_{t+1} \gets \sum_{k = 1}^{K} \frac{n_k}{n}w_{t+1}^k 
 \end{split}
 \label{eq:fedavg}
\end{equation}

\begin{figure*}[htb]
\centering
\includegraphics[width=\textwidth]{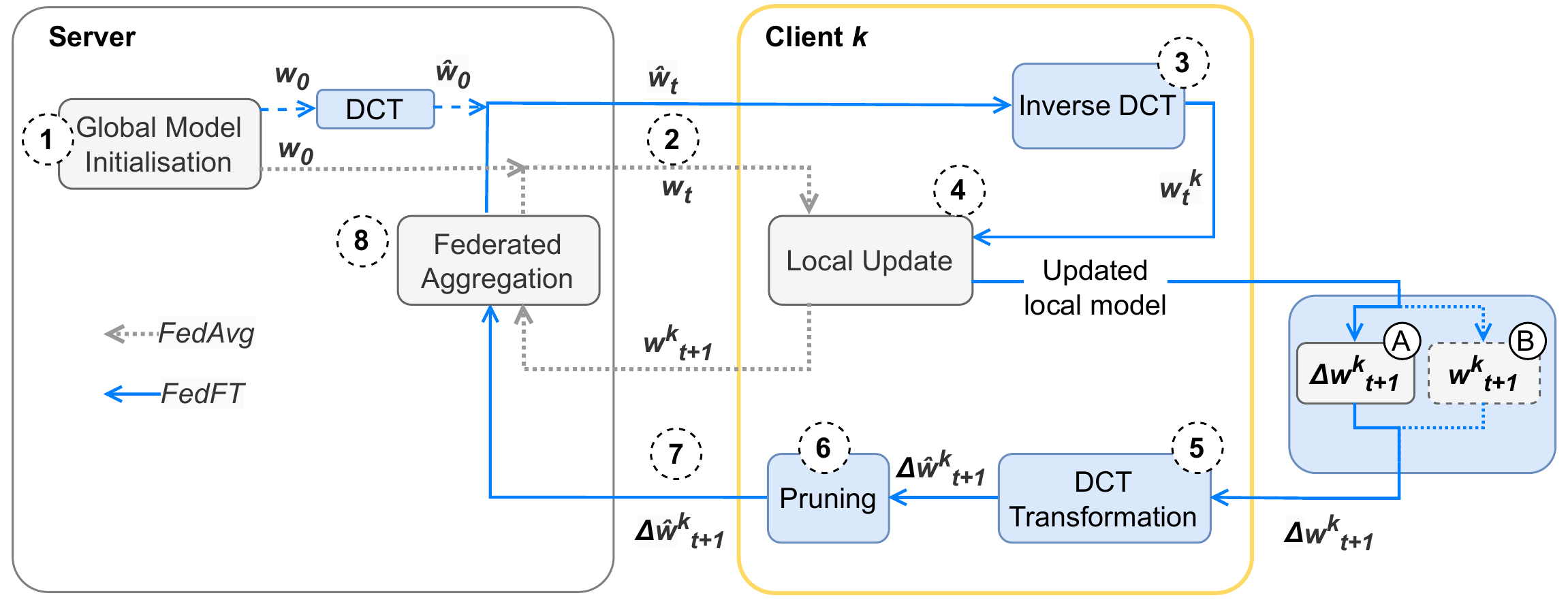}
\caption{Proposed \feddct methodology}
\label{fig:arch}
\end{figure*}

The aim of \feddct is to improve communication efficiency in FL through the utilisation of \fspace transformations on the model parameters. The overall \feddct, client and server communication setting is presented in Figure~\ref{fig:arch}. The original \fedavg phases are shown as grey-filled rectangles, i.e.,  
the initialisation, local update, and federated aggregation. 
Here, Steps 2 and 7 refer to downstream and upstream communications forms. Contributions of \feddct are the blue-filled rectangles. The blue and grey communication lines differentiate steps in relation to \feddct and \fedavg respectively.

\subsection{Global Model Initialisation}
The first step in Figure~\ref{fig:arch} is the initialisation of the global model, $w_0$, at $t=0$, which is common to both \feddct and \fedavg.
Additionally, \feddct converts, $w_0$, into \fspace using a transformation function, $T$, to obtain $\hat{w}_0$, which is communicated to all clients. 
\feddct can be applied even if the initial global model is pre-trained, such as a language model~\cite{tian2022fedbert} or transferred from another domain~\cite{florescu2022federated}, by converting the pre-trained weights into \fspace.

\subsection{Communication}
\label{sec:com}
The communication of model parameters in FL happens in two directions: from server to client~(downstream) and from client to server~(upstream). In both cases, \feddct communicates model parameters in \fspace using DCT-IV, a linear lossy function which is further discussed in Section~\ref{sec:fspace}.

\subsection{Client Local Update}
\label{sec:local}

Local update for a supervised task typically employs stochastic gradient descent~(SGD) over a number of epochs using local training data~(Step 4 in Figure~\ref{fig:arch}). 
In \fedavg this local update is applied to the model received through downstream communication from the server. 
With \feddct, the downstream communications of the initial and follow-on models, $w_0$ and $w_t$; are communicated in \fspace, as transformed models, $\hat{w}_0$ and $\hat{w}_t$; 
accordingly, an additional step of inverse transform, $\hat{T}(.)$, 
is required, 
where $\hat{T}(.)$ reconstructs the model parameters from \fspace to tensor space where local model updates can take place.
We acknowledge the possibility of performing these updates in \fspace, as referenced in~\cite{liu2018frequency}. 
However, we have chosen to maintain our approach, which helps to evaluate communication efficiency in isolation and enables us to assess \feddct on a diverse set of federated methodologies~(\fedavg, \fedprox and \fedsim) and neural models, all of which commonly operate in tensor space.
In our research, we examine two methods for representing locally updated models prior to transforming them into \fspace (using $T(.)$ for upstream communication to the server in Step 5 of Figure~\ref{fig:arch}. 
In the figure these alternative routes are labelled as (A) and (B) and refer to the following:

\subsubsection{Difference model (A)}
The purpose of this method is to capture only the net changes from local training, as the server can update the global model by adding these differences to its existing version, thus efficiently reconstructing the complete model.
Where updated local model parameters $w_{t+1}^k$ are compared against the received global model $w^k_t$ and the differences~($\Delta w_{t+1}^k = w_{t+1}^k - w^k_t$) are transformed into \fspace and communicated to the server. 
This is similar to the FL methodologies where client model update differences are communicated to the server~\cite{felix2020}, except we do so in \fspace.

\subsubsection{Complete model (B)}

If the objective is to conserve computational resources on the server when handling incoming updated models, opting to send the complete model is advantageous. However, this involves sending more parameters from each client which restricts the potential for compacting the models for efficient communication.
Where $w_{t+1}^k$ is transformed into \fspace and communicated to the server. This is simply the general FL methodology from~\cite{fedavg}.

We present the case for why $\Delta \hat{w}_{t+1}^k$ (difference model) is a more favourable choice compared to $\hat{w}_{t+1}^k$ (complete model) in Section~\ref{sec:fspace}.

\subsection{Pruning of Model Parameters}
\label{sec:pruningmodel}
Pruning allows FL to operate at varying levels of compression, thereby improving the efficiency of upstream communication. With \feddct, we have the option to implement pruning at Step 6 in Figure~\ref{fig:arch}, i.e. after performing the DCT transformation but before the upstream communication~(Step 7). The parameters pruned are the least significant coefficients of the updated client model in \fspace~(either $\hat{w}_{t+1}^k$ or $\Delta \hat{w}_{t+1}^k$ ). 
In the case of \feddct, pruning on DCT coefficients results in lossy compression where it approximates and discards some of the less significant frequency coefficients. 
Optimised compression with DCT is possible when a significant amount of the model parameters are captured within low-frequency coefficients.

Pruning then becomes an effective technique where the magnitudes of a specified percentage of high-frequency coefficients are set to 0 while minimising the reconstruction error. This is because, the high-frequency coefficients often correspond to features with high variance, i.e. noisy information. 
The pruning function and pruning percentage are referred to as  $P(.)$ and $\alpha$. 
Pruning can be applied once convergence is close, at which point most of the model will be contained within a low-variance. 
The implications of pruning in \fspace are discussed in Section~\ref{sec:fspace} with empirical findings in Section~\ref{sec:results}.

\begin{figure}[htb]
\centering
\includegraphics[width=0.85\columnwidth]{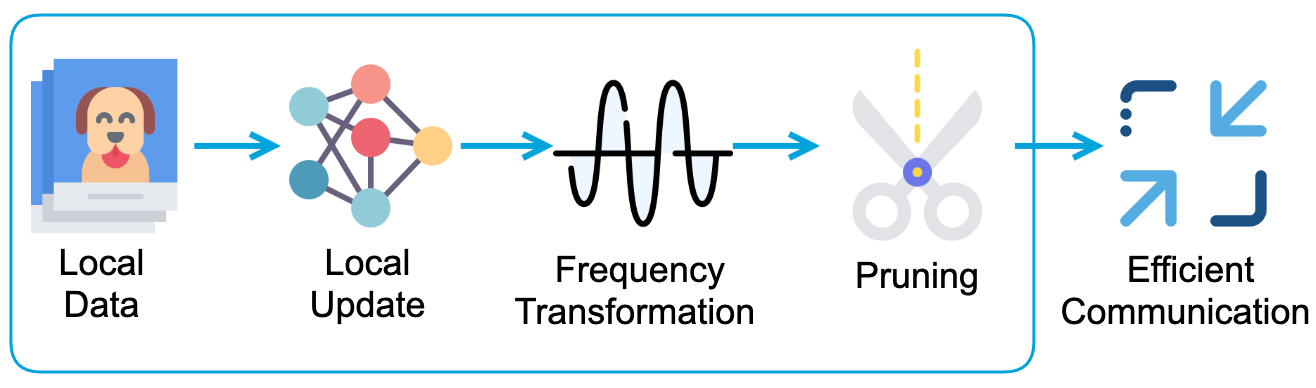}
\caption{Client level steps of \feddct}
\label{fig:clientprocess2}
\end{figure}

Figure \ref{fig:clientprocess2} provides a visual summary of the client-side steps involved in the \feddct algorithm. Following a local model update and transformation into \fspace, the model is pruned as detailed in this section. This process results in a condensed model that is then transmitted to the server, ensuring communication efficiency.

\subsection{Federated Aggregation}
A linear transformation function such as DCT-$IV$ is useful for performing federated aggregation in \fspace. 
If the transformation was non-linear this would require additional inverse transformations at the server to reconstruct the models in tensor space before federated aggregation can be performed and transformed thereafter for downstream communication. 
The use of DCT as the $T(.)$ function enables \feddct to carry out its aggregation in \fspace. 
It can do so with either the difference models~(see Equation~\ref{eq:agg-diff} with $\Delta \hat{w}_{t+1}^k$) or complete models~(see Equation~\ref{eq:agg-full} with $\hat{w}_{t+1}^k$) based on the selected approach for the local update step.

\begin{equation}
 \begin{split}
  \hat{w}_{t+1} \gets \sum_{k \in K} \frac{n_k}{n}(\hat{w}_{t}+\Delta \hat{w}_{t+1}^k )
\end{split}
 \label{eq:agg-diff}
\end{equation}
\begin{equation}
 \begin{split}
    \hat{w}_{t+1} \gets \sum_{k \in K} \frac{n_k}{n}\hat{w}_{t+1}^k 
 \end{split}
 \label{eq:agg-full}
\end{equation}
In Equation~\ref{eq:agg-diff}, the calculation of the weighted average includes the addition of the model changes to the previously maintained model on the server, which distinguishes it from the other (Equation~\ref{eq:agg-full}).

\subsection{\feddct Algorithm}
Algorithm~\ref{algo:feddct} brings together the extensions proposed with \feddct.
Line 1 performs the initial global model transformation into \fspace, once received by clients each performs the inverse transformation in Line 6, prior to carrying out the local update. 
Once completed, the client calculates the $\Delta w^k_{t+1}$~(Line 8), and performs \fspace transformation and pruning with the percentage of pruning controlled by $\alpha$~(Line 9). 
Once the client models in \fspace $\Delta \hat{w}^k_{t+1}$ are communicated to the server, it performs federated aggregation on the updated local models in \fspace.

In the algorithm areas highlighted in blue text signify the specific modifications we have implemented to adapt our proposed method to the vanilla \fedavg methodology.

\begin{algorithm}
\caption{FedFT}
\label{algo:feddct}
\begin{algorithmic}[1]
\REQUIRE $w_0$: initialised global model, \textcolor{blue!80}{$\alpha$: Pruning Rate}, $K$: number of selected clients per round
\REQUIRE \textcolor{blue!80}{$T(.)$ DCT Function, $\hat{T}(.) $ Inverse DCT Function, $P(.)$ Pruning Function}
\STATE \textcolor{blue!80}{${\hat{w_0}} =  T(w_0) \gets$ DCT transformation}
\FOR{t=0,1,2, ...}
\STATE Broadcast $\hat{w}_{t}$ to all clients
\STATE Select $K$ clients
    \FORALL{$k \in K$}  
        \STATE  \textcolor{blue!80}{$w_t^k = \hat{T}(\hat{w}_t) \gets$ inverse DCT transform}
        \STATE $w^k_{t+1} \gets$ update $w^k_t$ using SGD on client data
        \STATE \textcolor{blue!80}{$\Delta w_{t+1}^k = w_{t+1}^k - w^k_t \gets $ update differences}
        \STATE \textcolor{blue!80}{$\Delta \hat{w}_{t+1}^k = P(T(\Delta w_{t+1}^k), \alpha) \gets $ DCT transform and prune}
        \STATE \textcolor{blue!80}{Send $\Delta \hat{w}_{t+1}^k$ to the server}
    \ENDFOR
    \STATE $\hat{w}_{t+1} \gets \sum_{k \in K} \frac{n_k}{n}(\hat{w}_{t}+ \Delta \hat{w}_{t+1}^k ) \gets$  Federated Aggregation on update differences
\ENDFOR
\end{algorithmic}
\end{algorithm}

\section{Role of Model Variance for transformed communication}
\label{sec:fspace}

Based on an analysis of literature~(see Section~\ref{sec:background}), we select DCT as the transformation technique to convert $w$ into \fspace. Where a given set of model parameters, $w$ is a multi-dimensional array~(i.e. a tensor) where the number of dimensions depends on the model architecture. 
Out of the DCT variants, DCT-$IV$ is selected due to its linear, orthogonal and symmetric properties required for inverse transformations and necessary for federated aggregation. 

Equation~\ref{eq:dct-2d} presents the DCT-$IV$ transformation function $T(.)$ for $w$ represented in a tensor space of 
$\mathbb{R}^{N\times M}$,  
where $k \in \{0, \ldots, N-1\}$ and $l \in \{0, \ldots, M-1\}$ respectively.
%$w \in \mathbb{R}^{N\times M}$. 
%; while its multi-dimensional representation in the tensor-space conveys the generalisability of the transform function.
\begin{equation}
% \small
\begin{gathered}
\hat{w}_{k,l} =
 \sum_{n=0}^{N-1}\sum_{m=0}^{M-1} w_{n,m} \cos \left(\frac{\pi (2m +1)(2k+1)}{4N}\right) \cos \left(\frac{\pi (2n +1)(2l+1)}{4M}\right)\\
%\forall \quad k = 0,\ \ldots\ N-1 ~.\\
%\forall \quad l = 0,\ \ldots\ M-1 ~.\\
\end{gathered}
\label{eq:dct-2d}
\end{equation}
Without loss of generalisability, $w$ represents a set of model parameters between two fully connected layers of a neural architecture. 
With multi-dimensional tensors, beyond just 2-dimensions, the summations can be extended over the additional dimensions.

Equation~\ref{eq:dctin-2d} is the inverse function $\hat{T}(.)$, where $n \in \{0, \ldots, N-1\}$ and $m \in \{0, \ldots, M-1\}$. 
\begin{equation}
% \small
\begin{gathered}
w'_{n,m} = \frac{2}{N}
 \sum_{k=0}^{N-1}\sum_{l=0}^{M-1} \hat{w}_{k,l} \cos \left(\frac{\pi (2m +1)(2k+1)}{4N}\right) \cos \left(\frac{\pi (2n +1)(2l+1)}{4M}\right)\\
%\forall \quad n = 0,\ \ldots\ N-1 ~.\\
%\forall \quad m = 0,\ \ldots\ M-1 ~.\\
\end{gathered}
\label{eq:dctin-2d}
\end{equation}
Accordingly, the reconstruction loss is calculated as $|\hat{T}(T(w)) - w|$. 

\begin{figure}[t]
  \centering
    \includegraphics[width=0.75\textwidth]{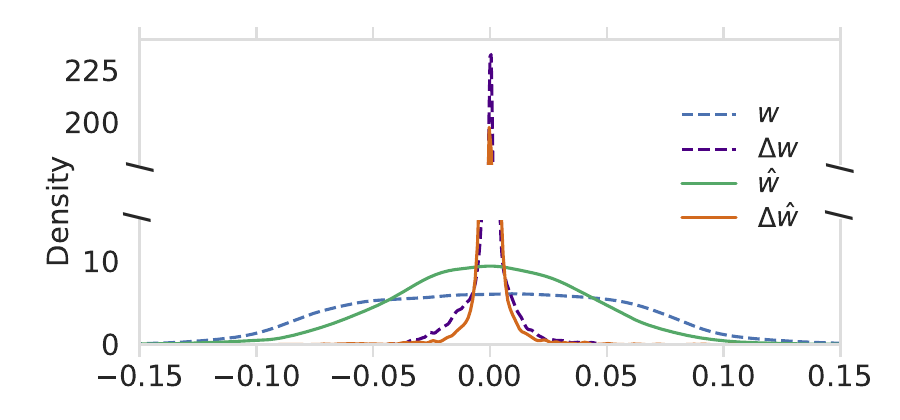}
    \caption{Density in tensor and frequency spaces}
    \label{fig:real-w-delta-w}
\end{figure}

The distribution of the tensor space directly impacts the magnitude of the DCT coefficients and the way they are distributed. This in turn affects the level of pruning possible to manage reconstruction error after the inverse transform~\cite{lam2000mathematical}. 
We observe the distribution of the tensor space conforms to a Gaussian distribution which can be expressed using mean and variance (Figure \ref{fig:real-w-delta-w}). 
Accordingly, the variance of model parameters, $w^k \in \mathbb{R}^{N\times M}$, for any given round is calculated as in Equation~\ref{eq:var-complete}, where $\bar{w}^k$ indicates the mean of model parameters. 
\begin{equation}
 \begin{gathered}
Var(w^k) = \frac{1}{N\times M}\sum_{n=0}^{N-1}\sum_{m=0}^{M-1} (w^k_{n,m} - \bar{w}^k)^2\\
 \end{gathered}
 \label{eq:var-complete}
\end{equation}
Similarly, the variance of the difference model, $\Delta w^k \in \mathbb{R}^{N\times M}$, can be calculated as in  Equation~\ref{eq:var-diff}.% where $\Delta \bar{w}^k_{t+1}$ is the mean.
\begin{equation}
 \begin{gathered}
Var(\Delta w^k) = \frac{1}{N\times M}\sum_{n=0}^{N-1}\sum_{m=0}^{M-1} (\Delta w^k_{n,m} - \bar{\Delta w}^k)^2
 \end{gathered}
 \label{eq:var-diff}
\end{equation}

% UPDATED: ADDED
We carried out an empirical study to better understand these distributional relationships between tensors and how it transforms into \fspace in the context of variance. The following observations were made: the variance of $Var(\Delta w^k)$ remains consistently below that of $Var(w^k)$ throughout the communication rounds; tensor space~(for both $w$ and $\Delta{w}$) conform to a Gaussian distribution (Figure~\ref{fig:real-w-delta-w}); Further the corresponding frequency space in the form of DCT coefficients~(for both $\hat{w}$ and $\Delta{\hat{w}}$) also conform to a Gaussian distribution (Figure~\ref{fig:real-w-delta-w}); frequency space has lower variances, compared to tensor space under the strict constraint that $w$ is a set of model parameters that are optimised using SGD.
Accordingly, at any given round, it is reasonable to assume that the inequalities between the variances in the tensor space are also likely to hold in the frequency space~(Equation~\ref{eq:var-dctvar}).

\begin{equation}
\begin{gathered}
Var(\Delta w^k) < Var(w^k) \Longleftrightarrow Var(\Delta \hat{w}^k) < Var(\hat{w}^k) 
\end{gathered}
\label{eq:var-dctvar}
\end{equation}

\begin{figure}[htb]
\centering
\includegraphics[width=\columnwidth]{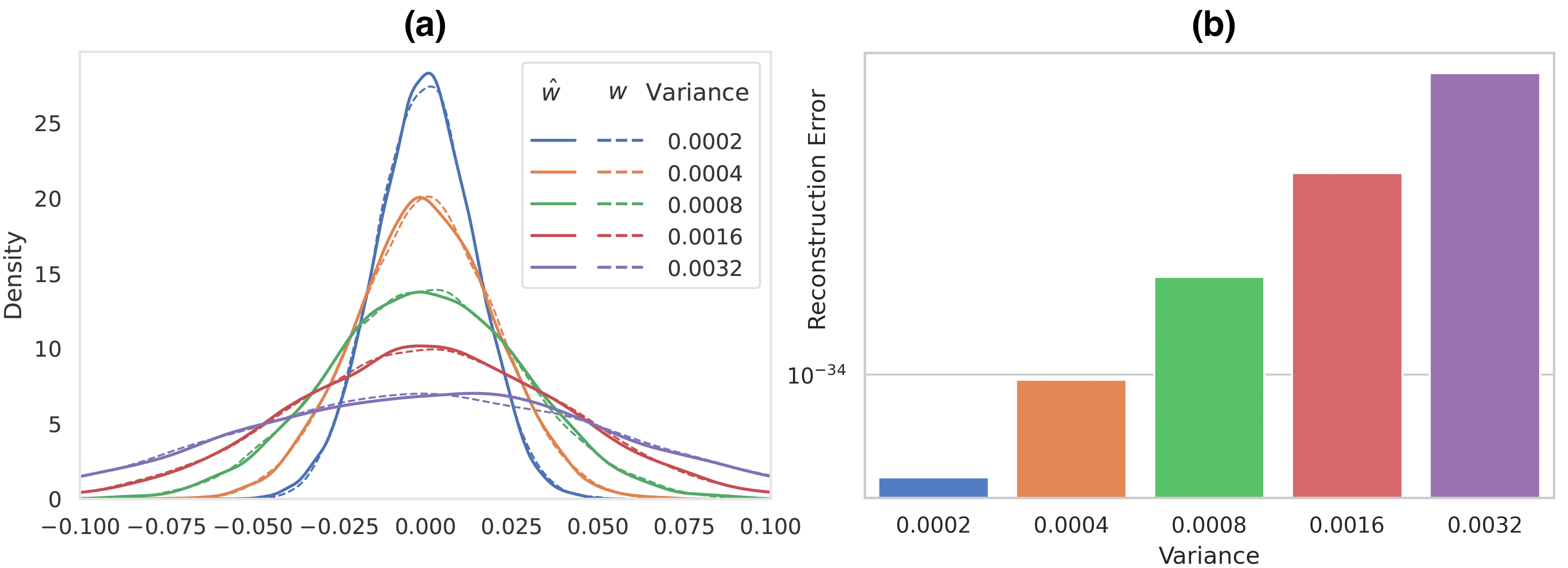}
\caption{Variance and reconstruction error relationship}
\label{fig:gaussian}
\end{figure}

In Figure \ref{fig:gaussian} we study the link between variance and reconstruction error. The plots show five synthetic Gaussian distributions, each with 0 mean and 10,000 samples for increasing variances (a) and their corresponding reconstruction errors (b), the x-axis is the variance, and the y-axis is the reconstruction error. It is clear from these plots that there is a direct relationship between increasing variance in distributions and increasing reconstruction errors. This confirms the benefits of using the difference model over the complete model and highlights the advantages of reduced variance in the frequency space for optimising compression in communication. 

\begin{figure}[htb]
    \centering
    \includegraphics[width=.65\textwidth]{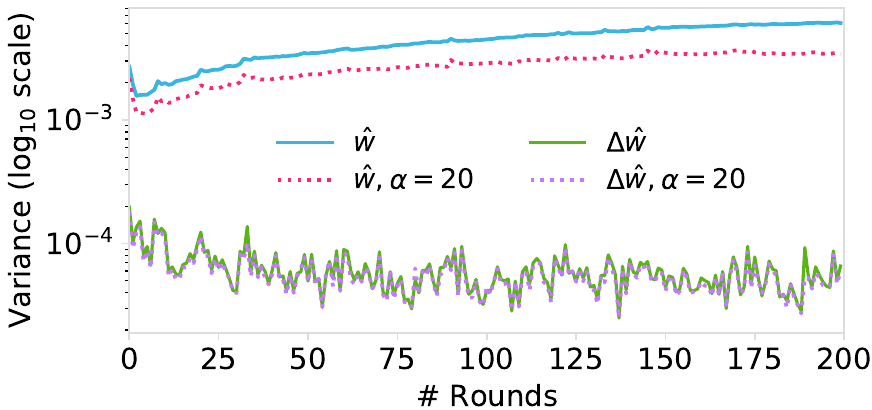}
    \caption{Variance in frequency space \& pruning}
    \label{fig:variance-w-deltaw-pruned}
\end{figure}

Finally in Figure~\ref{fig:variance-w-deltaw-pruned} we analyse how pruning affects model parameters in \fspace. 
Here, the variances in the y-axis are on a log scale and the x-axis is on communication rounds. 
This plot further verifies the assertion made in Equation~\ref{eq:var-dctvar} that in \fspace, the variance of the difference model is less than that of the complete model~(blue and green lines). 
We use variance here as a proxy for reconstruction error, where increasing variance~(and so increasing reconstruction error) indicates the diminishing utility of pruning. 

Accordingly, we use a pruning rate, $\alpha=20\%$, to study the impact of pruning less significant coefficients in the frequency space. 
We can see that pruning the difference~($\Delta \hat{w}$) results in hardly any drop in variance. In contrast, a noticeable drop in variance is observed when using the complete model~($\hat{w}$).
We can conclude from these empirical observations that utilising the difference model, $\Delta \hat{w}$, for \feddct will yield better results as compared to using the complete model, as stated in Equation~\ref{eq:agg-diff} vs Equation~\ref{eq:agg-full}. We will be using this version of \feddct in our comparative study.

\section{Experiment Setup}\label{sec:eval}
We evaluate the performance of \feddct, with respect to three important aspects. First, its generalisability to existing FL baseline methodologies. Second, we investigate its applicability to various complex neural architectures. Finally, we analyse the impact of pruning with \feddct on performance and communication efficiency.

\subsection{Datasets}
\label{sec:datasets}
The generalisability of \feddct is evaluated with four real-world datasets consisting of two image datasets, one time-series dataset and one text dataset all of which perform multi-class classification.
We reuse the MNIST and FEMNIST datasets, which were originally introduced in the work of \cite{fedprox}. Additionally, we utilise the \greads and \mex datasets, which were initially introduced in the work of \cite{fedsim}. To ensure compatibility with a realistic non-IID (non-independent and identically distributed) setting, all the datasets used in our study enforce statistical heterogeneity by restricting the number of classes per client. This approach guarantees that the proposed \feddct methodology accommodates diverse and realistic scenarios.

\begin{itemize}
    \item \textbf{MNIST} is a handwritten digit recognition dataset adapted to the FL setting as proposed in~\cite{fedprox}. The dataset contains 69,035 data instances of hand written digits of 10 classes distributed among 1000 clients and each client has samples for only 2 classes. A data instance is an image of size 28$\times$28. 
    \item \textbf{FEMNIST} (Federated-Extended-MNIST) is a handwritten character recognition dataset from~\cite{leaf}. 
    The subsample consists of 10 lowercase characters (a-j) and is used for a 10-class character classification task. This dataset is distributed among 200 clients, with each client having samples for only 3 classes. Each data instance in this dataset is an image with dimensions of 28x28.
    \item \textbf{\mex} is adapted to the FL setting in~\cite{fedsim} from MEx which is an exercise recognition dataset collected with 30 subjects performing 7 different physiotherapy exercises~\cite{mex}. The dataset has 934 data instances from a pressure mat where an instance is a sequence of heat maps~(size $5 \time 16 \time 16$) recorded for 5 seconds at 1Hz. Each client has a random amount of samples for only 2 exercise classes. 
    \item \textbf{\greads} is extracted from the book review dataset Goodreads and transformed to the FL setting in~\cite{fedsim}. It contains 100 clients each with 2-10 of their own reviews which emulates a heterogeneous FL setting. The task is to predict if a text review contains a spoiler or not.
\end{itemize}

\subsection{Baselines}
Selected from widely-accepted FL methodologies, excluding those specific to an application, dataset or model type:
\begin{itemize}
    \item \fedavg: general FL methodology from~\cite{fedavg}.
    \item \fedprox: variant of \fedavg focused on improving stability and performance in non-IID settings using regularisation in client update from~\cite{fedprox}.
    \item \fedsim variant of \fedavg  that performs clustered federated aggregation based on latent similarity knowledge between clients~\cite{fedsim}. 
\end{itemize}

Similar to the approach in Algorithm \ref{algo:feddct}, where \feddct was implemented with \fedavg, the adaptations of \feddct for \fedsim and \fedprox are described in \ref{ap:algos}. Algorithms \ref{algo:ftfedsim} and \ref{algo:ftfedprox} detail the application of \feddct within the \fedsim and \fedprox methods, respectively.
 
\subsection{Summary of Experiments}

In Table \ref{tab:exsummary}, we present a comprehensive summary of the experiments carried out in this study. Detailed descriptions of each experiment are provided in the corresponding subsections. The table showcases the range of datasets, baseline methodologies and neural architectures employed in our experiments. 
\begin{table}[ht!]
\centering
\caption{ Comprehensive summary of \feddct experiments across diverse datasets, baselines and model architectures}
\small
\renewcommand{\arraystretch}{1.2}
\begin{tabular}{p{40mm}p{50mm}p{32mm}}
\toprule
\textbf{Experiment}  &  \textbf{Objective} &  \textbf{Setup}  \\
\midrule
Comparing frequency transformation methods  & To select which frequency transformation method is suitable & Baseline: \fedavg \newline Datasets: MNIST \newline Model: MLR \\
\hline
Comparison of different variants of DCT  & To select which variant of DCT is most applicable  & Baseline: \fedavg \newline Datasets: MNIST \newline Model: MLR \\
\hline
Generalisability of \feddct  & Study the applicability of \feddct & Datasets: All \newline Model: MLR \\
\hline
Evaluation with different learning models & Study generalisability with different neural architectures   & Datasets: All \newline Models: CNN-2D, MLP-3 and RNN  \\
\hline
Evaluation of compute resources & Effect on computation overheads & Datasets: All  \newline Model: MLR \\
\hline
Analysing statistical heterogeneity & Investigate the impact on \feddct with varying levels of statistical heterogeneity & Datasets: FEMNIST(1-3) \newline Model: MLR \\
\hline
Impact of \feddct pruning and communication efficiency  & Investigate the communication performance of \feddct  & Datasets: All \newline Model: MLR \\
\hline

Impact of \feddct pruning on \fedsim   & Investigate the impact of using \feddct on the \fedsim method  & Datasets: All \newline Model: MLR \\
\hline

Impact of pruning post-convergence  & Investigate the impact of using \feddct as a post-convergence method  & Datasets: MNIST and \mex \newline Model: MLR \\
\bottomrule
\end{tabular}
\label{tab:exsummary}
\end{table}
\feddct and baseline methodologies were implemented using 
Python with Tensorflow \cite{tensorflow}
libraries extending the setup from \fedprox and the source code is available on GitHub\footnote{https://github.com/chamathpali/FedFT}.

\subsection{Generalisability of \feddct}
\label{sec:compare-meds-eval}

To study the generalisability of \feddct,
we adapted the baseline FL methodologies to \feddct and compared against their original form.
The comparison of FL methodologies adapted for \feddct is carried out with a Multinomial Logistic Regression~(MLR) model trained for classification.
We conduct a comprehensive evaluation using all of the selected baselines and datasets to represent a broad range of scenarios in FL settings.
By integrating \feddct into different baseline methodologies, we could observe its performance and efficacy in comparison with their original forms.

% UPDATED - REMOVED THIS SINCE THE ABOVE LINE IS THERE
% Local update hyper-parameters are as follows: number of epochs is 20; batch size is 10; learning rates are 0.03, 0.003, 0.01 and 0.3 for MNIST, FEMNIST, \mex and \greads; and pruning rate is $0\%$. 
% The number of communication rounds are limited to either after convergence or maximum of 1000 rounds; accordingly, corresponding communications rounds~(and clients per round) for MNIST, FEMNIST, \mex and \greads were 200(20), 500(20), 200(10) and 500(20). 
% Each client dataset is split for train and test at 80\% and 20\%. 
% The cluster sizes for \fedsim are reused from~\cite{fedsim} as 5, 9, 3 and 11 for MNIST, FEMNIST, \mex and \greads.

\subsection{Generalisability to Neural Architectures}
\label{sec:compare-arch-eval}
\feddct algorithm transforms model parameters to \fspace using multi-dimensional DCT. Accordingly, applicability of \feddct to different neural architectures that are of different dimensions is key to generalisability. We evaluate this with the three most commonly used neural architectures: Multi-layer Perceptrons~(MLP); Convolutional Neural Networks~(CNN); and Recurrent Neural Networks~(RNN). The dimensions of the model parameters are summarised in Table~\ref{tbl:compare-archs}. 

\begin{table*}[htb]
\caption{Alternative neural architectures}
\label{tbl:compare-archs} 
\centering
\small
\footnotesize
\begin{tabular}{p{0.1\textwidth}p{0.1\textwidth}p{0.30\textwidth}p{0.25\textwidth}p{0.08\textwidth}}

\toprule
Dataset& Model & Architecture&$|w|$ & Params\\
\midrule
FEMNIST MNIST & CNN-2D & $conv2d(5,5)64 \rightarrow maxpool(2,2) \rightarrow conv2d(5,5)64 \rightarrow $ $ maxpool(2,2) \rightarrow dense(2048) \rightarrow dense(10)$  &$[[5,5,32]$, $[32]$, $[5,5,64,32]$, $[64]$,$[3136,2048]$, $[2048]$, $[2048,10]$, $[10]]$ & 6.49M \\
\midrule
\mex & MLP-3& $ dense(1280) \rightarrow dense(640)\rightarrow$&$[[1280,1280]$, $[1280]$,$[1280,640]$,$[640]$,\\
&&$dense(120)\rightarrow dense(7)$&$[640,120]$, $[120]$,$[120,7]$,$[7]]$  & 2.53M \\
\midrule
\greads & RNN & $embedding(25) \rightarrow rnn(128)\rightarrow dense(2)$ & $[[25,25]$, $[25,128]$, $[128,128]$, \textit{$[128]$}, $[25,128]$, $[128,2]$,$[2]]$ & 20K\\
\bottomrule
\end{tabular}
\end{table*}

It is important to highlight that each architecture makes use of a unique multi-dimensional tensor. For the FEMNIST and MNIST datasets, a CNN-2D architecture with 6.49 million parameters is employed. In the case of \mex, a deep neural network called MLP-3 is utilised with 2.53 million parameters. Lastly, the \greads dataset employs an RNN architecture. 
The hyperparameters are kept same with the exception of reducing the number of local epochs to 10 and the learning rate to 0.0001, to prevent over-fitting on the \greads dataset. Note that these experiments are conducted with a pruning rate of $\alpha=0\%$.

\subsection{Impact of \feddct Pruning}
\label{sec:compare-prune-eval}

\feddct applies pruning to improve communication efficiency which is lossy and can impact overall performance. Accordingly, we explore the performance impact of pruning with MLR models trained on four datasets with increasing $\alpha$ rates. 
We explore two variants of pruning: one applied from the start of communication~(round=0) and the other applied after the model has converged~(round$\sim$50).
In each case, we compare different pruning rates where $\alpha$ is varied from 0\%~(no pruning) to $\sim50\%$ in increments of $\sim10\%$. The actual percentages for MLR models depend on the output layer size; for example on \mex, where $|\hat{w}|=[1280,7]$, $\alpha=\sim14\%,\sim29\%,\sim43\%~ and \sim57\%$ for when 1,2,3, and 4 weights are set to 0 in each of 7 weights.  
Each experiment plots the test accuracy over communication rounds. Furthermore, we evaluate how pruning impacts communication efficiency by plotting the cumulative communication cost in MegaBytes (MB) over 200 rounds for each dataset. This is repeated for all values of $\alpha$ to determine the optimal value that can maintain test accuracy (as close to accuracy with no pruning i.e., when $\alpha=0$) while minimising the cost in MB. 

\subsection{Analysing the Impact of non-IID on \feddct} 

This experiment focuses on evaluating the influence of non-IIDness on the effectiveness of the proposed \feddct method. 
The datasets utilised in the \feddct experiments are carefully selected to reflect their realistic non-IID nature. These datasets are chosen based on previous research in FL, as discussed in Section \ref{sec:datasets}. 
In this analysis, we employ the FEMNIST dataset as our core dataset. To evaluate the impact of \feddct across varying degrees of non-IID, we purpose three variants of the FEMNIST dataset: FEMNIST(1) with one class per client, FEMNIST(2) with two classes per client, and FEMNIST(3) with three classes per client.
The default configuration of the FEMNIST dataset used for primary experiments typically consists of three classes per client.

\subsection{Performance metrics}
\label{sec:metrics}
In selecting all hyper-parameters, we prioritised ensuring comparability and reproducibility with~\cite{fedprox} and~\cite{fedsim}. 
Hyper-parameter details are summarised in Table~\ref{tbl:exdetails}.
The primary performance metric is the test accuracy of the global model against individual client test data, adapting the evaluation setup from~\cite{fedprox}.
The test accuracy at any given round is the mean of all test client accuracy values weighted by their test set sizes. 
For generalisability and to reduce the sampling error, each experiment is repeated 35 times with different random seeds. 
Results plot mean test accuracy at a given communication round calculated as the mean over the 35 trials. 
We introduce a secondary metric, denoted as $\Theta(.)$, to estimate the communication cost of a model $w$ in MBs.
We measure the upstream communication cost accumulated over $t$ communication rounds per client as $t \times \Theta(P(T(w),\alpha))$.

\begin{table}[htb]
\caption{Hyper-parameter details}
\centering
\small
\renewcommand{\arraystretch}{1.1}
\begin{tabular}{lrrrrrr}
\toprule
 & & Learning & Total & Clients per & Number of \\
 Dataset  & Features  &    rate  & clients & round &  clusters   \\
\midrule
MNIST & 784 & 0.03 & 1,000  &  20 & 5\\
FEMNIST & 784 & 0.003 & 200  &  20 & 9\\
\mex & 1280 & 0.01 & 30 & 10  & 3 \\
\greads  & 2517 & 0.3 & 100  &  20 & 11\\
\bottomrule
\end{tabular}
\label{tbl:exdetails} 
\end{table}

\section{Results and Discussion}
\label{sec:results}
In this section, we conduct a detailed analysis of the experimental results and provide a discussion of the findings.
\subsection{Comparing Frequency Transformation Methods}

We aim to optimise the function $T(.)$ for efficient communication in FL. To do this, we compare two well-known frequency transformation methods: DCT and FFT.
These methods are key for transforming model parameters into frequency space for \feddct as discussed in Section \ref{sec:methods}.
We designed an experiment to test how well DCT, particularly DCT-$IV$, works compared to FFT in FL settings. Our comparison looks at important factors for FL, including compression efficiency, information retention, and impact on the convergence rate of the learning process. 
We conducted this experiment on the MNIST dataset, applying the \fedavg baseline across 200 communication rounds. To ensure statistical robustness, we averaged the results over 35 separate runs, each initialised with a unique random seed.

\begin{figure}[htb]
\centering
\includegraphics[width=0.8\textwidth]{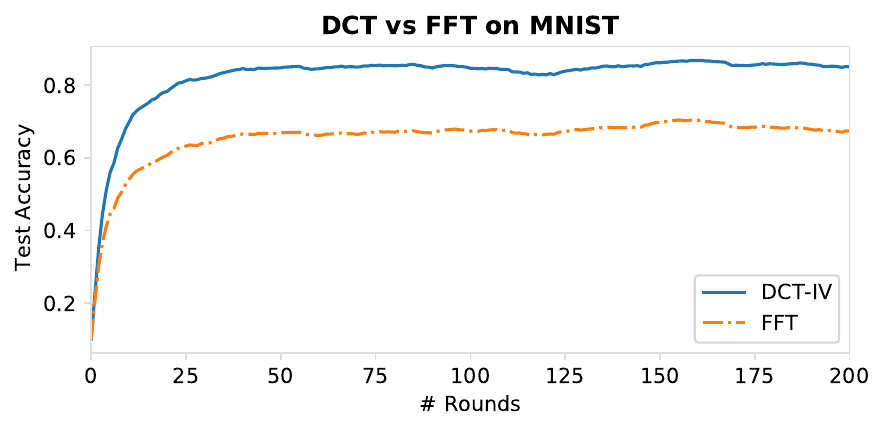}
\caption{Comparison of DCT and FFT on MNIST dataset}
\label{fig:compare-dct-fft}
\end{figure}

As illustrated in Figure \ref{fig:compare-dct-fft}, our results demonstrate a notable performance differential between the two methodologies. DCT-$IV$ emerges as a superior choice, offering significant advantages over FFT.
These advantages are quantified in terms of reduced communication overhead and enhanced model accuracy post-transformation. The superiority of DCT-$IV$ can be attributed to its inherent properties that align well with the sparsity and locality of model parameters in FL scenarios.

\subsection{Comparison of Different Variants of DCT}

\begin{figure*}[htb]
\centering
\includegraphics[width=\textwidth]{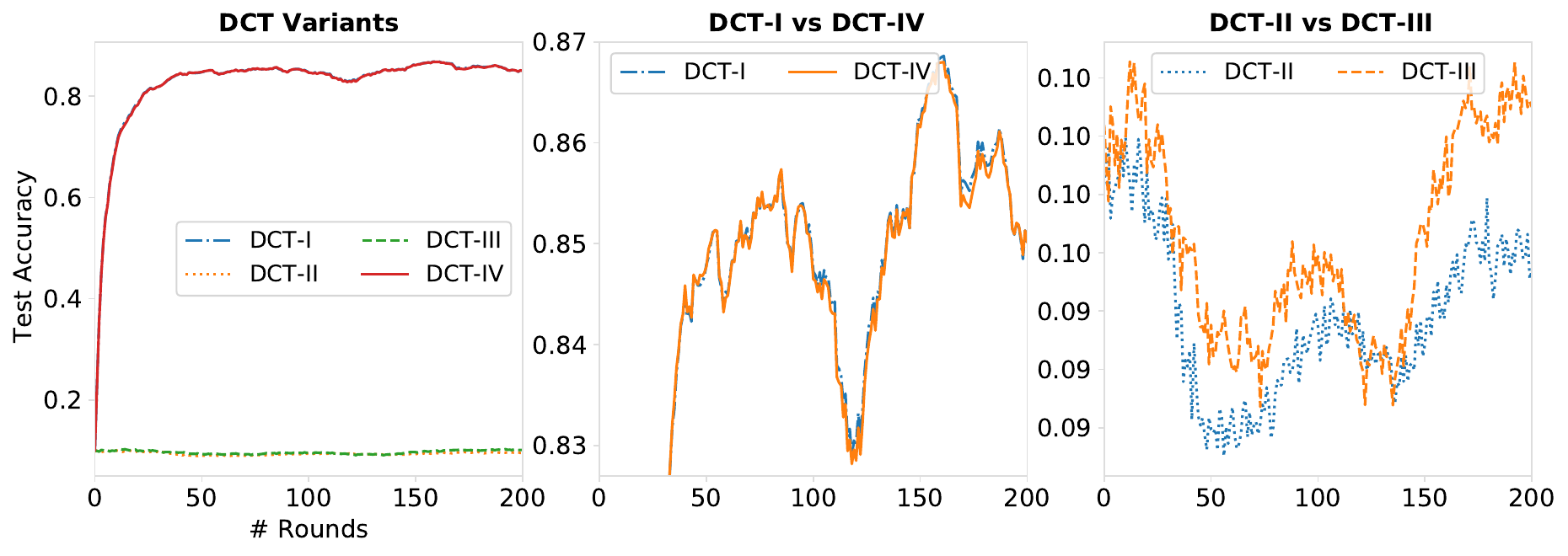}
\caption{Comparison of DCT variants (I to IV) on MNIST with \feddct with \fedavg}
\label{fig:compare-dct-types}
\end{figure*}

Evaluating the impact of various DCT variants is crucial as each variant has distinct characteristics and applications.
This experiment is designed to discover which DCT variant is most suitable for our specific needs with FL.
In Figure \ref{fig:compare-dct-types}, we present a comparative analysis of four DCT variants, identified as DCT-$I$ through DCT-$IV$. 

This experiment was conducted using the MNIST dataset, comparing the \fedavg baseline with our proposed \feddct algorithm across 200 communication rounds, averaging the results across 35 unique runs.
This comparison is crucial in understanding how each variant handles the transformation and compression of model parameters.
The results reveal a notable divergence in performance among these variants. Specifically, DCT-$I$ and DCT-$IV$ stand out for their efficiency, with lower reconstruction errors and indicating an accurate representation of the original model parameters.
In contrast, DCT-$II$ and DCT-$III$, while effective in their respective applications, show less favourable results in our context. Their performance, characterised by higher reconstruction errors which suggests not suitable to handle FL model parameters.

We have selected the DCT-$IV$ variant for our transformation function $T(.)$, primarily due to its lower computational demands and proficiency in managing large data structures. This makes DCT-$IV$ particularly well-suited for the diverse and computationally varied landscape of FL applications. 
All subsequent experiments in this study will utilise DCT-$IV$ as the transformation function.

\subsection{Generalisability of \feddct} 
\label{sec:generalisability}

\begin{figure*}[htb]
\centering
\includegraphics[width=\textwidth]{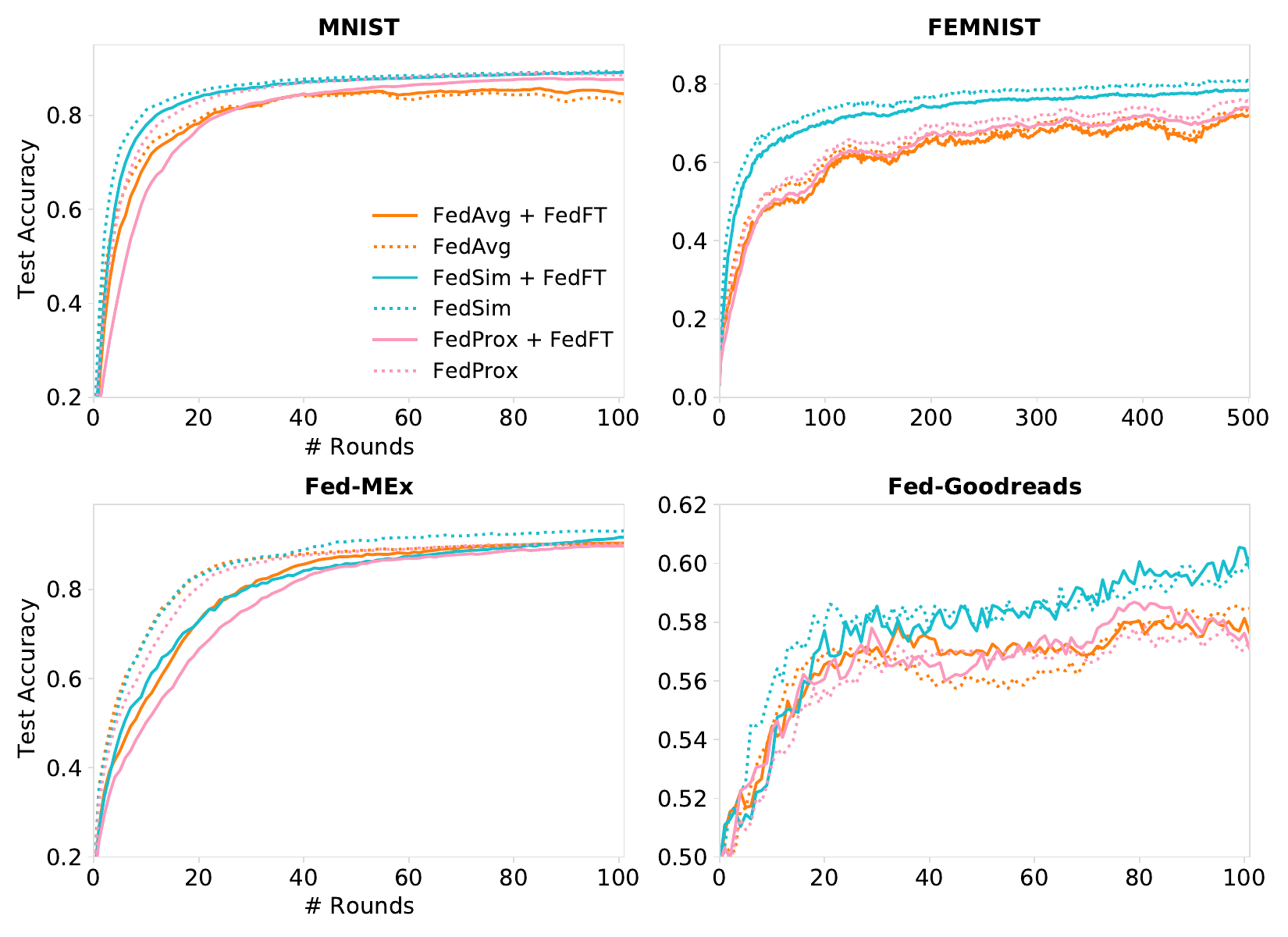}
\caption{Comparison of \feddct with baselines FL methodologies}
\label{fig:resultsmain}
\end{figure*}

Our primary experiments focus on assessing the generalisability of the proposed \feddct method. Following the setup outlined in Section \ref{sec:compare-meds-eval}, we evaluate the efficacy of \feddct across four datasets, comparing it with three state-of-the-art FL baselines.
Figure~\ref{fig:resultsmain} presents test accuracy results for increasing communication rounds with three FL methodologies, both with \feddct (solid line) and without \feddct (dotted line), across four datasets. Overall, \feddct adaptations match the performance of baseline counterparts at convergence, demonstrating that efficient communication of model parameters in frequency space does not compromise performance. 
The noticeable performance difference in the rounds prior to convergence across all methodologies on the \mex dataset is attributed to the small number of participating clients and their data sizes~(30 total clients and 10 selected per round). With fewer clients who have fewer samples, each local update makes larger weight adjustments~(high variance) resulting in significant changes to the global model.

As discussed in Section~\ref{sec:fspace}, high variance results in high reconstruction error and evidently affects the model performance before convergence.
The only significant performance loss post-convergence is observed with \fedprox on \greads dataset where \feddct adaptation of \fedprox fails to converge. 
We attribute this to the MLR classifier not being a suitable architecture; we recover this performance loss when using a recurrent neural model better suited to textual content is presented in Section~\ref{sec:arch-compare}.
This study not only demonstrates the practicality of integrating \feddct into various FL methodologies but also highlights its minimal impact on overall performance. 
This finding is significant as it highlight the adaptability and compatibility of \feddct with a wide array of FL methodologies and datasets.
The results suggest that \feddct could be a valuable tool in improving communication efficiency and privacy in FL systems, offering a balance between efficiency and performance.

\subsection{\feddct with Different Neural Architectures}
\label{sec:arch-compare}

\begin{figure*}[htb]
\centering
\includegraphics[width=\textwidth]{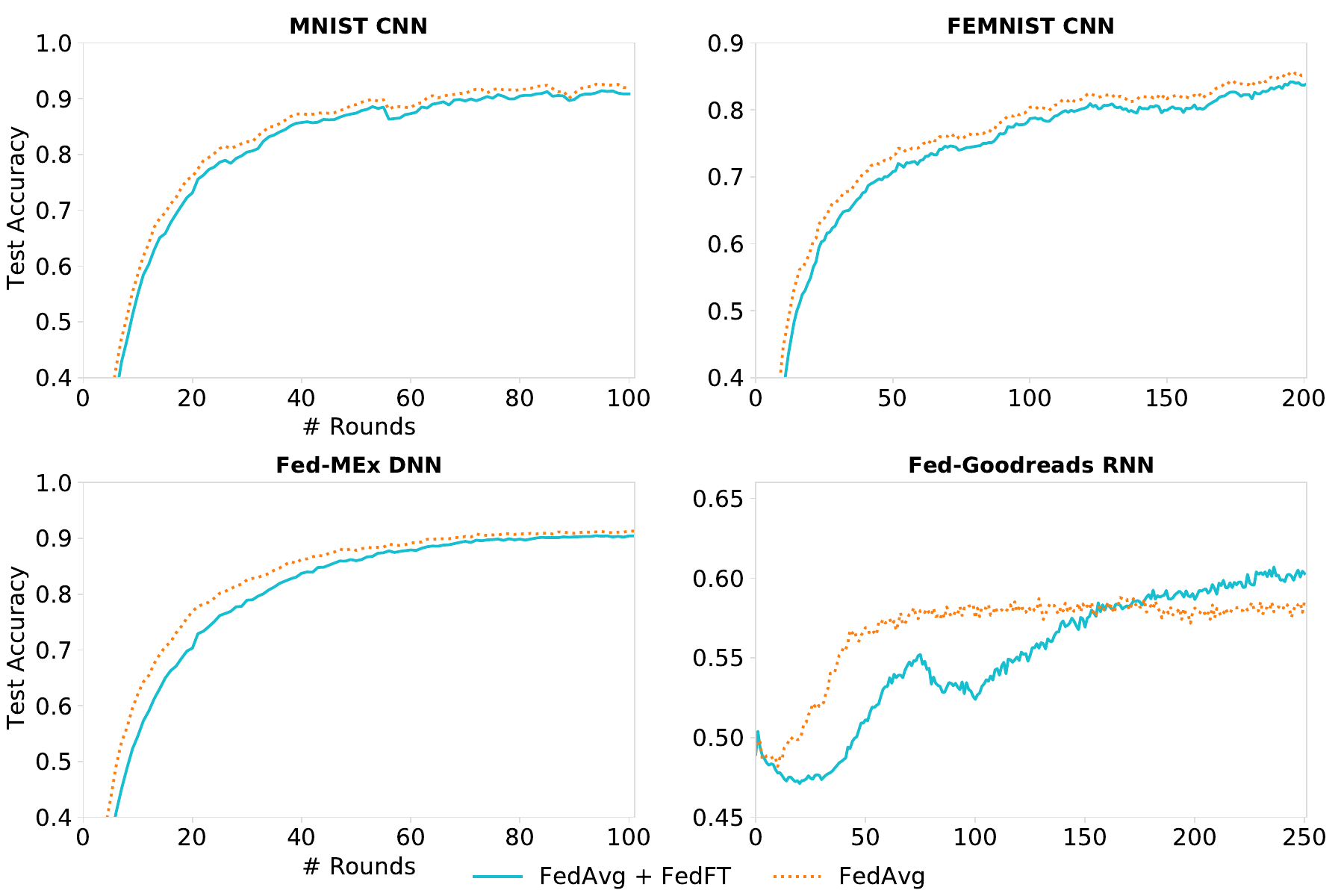}
\caption{\feddct using different neural architectures}
\label{fig:compare-archs}
\end{figure*}

To further understand the adaptability of \feddct, we next explore its impact on different neural architectures.
A summary of the architectures and the details of the experiment setup was described in Table~\ref{tbl:compare-archs}. 
In Figure~\ref{fig:compare-archs}, we present a comparative analysis of \feddct and \fedavg when applied on different neural architectures.

Overall, \feddct demonstrates comparable or superior performance compared to \fedavg.
For instance, the CNN model trained on MNIST has approximately 6.5 million parameters that are transformed between the tensor space and \fspace at each communication round without affecting overall performance. Similar performance is observed with \mex which trains a DNN architecture.
The RNN model trained on \greads with \feddct shows a drop in performance in early communication rounds, however, it improves and surpasses \fedavg performance after round $\sim 150$. We attribute this improved performance to the imperceptible reconstruction error in DCT-$IV$ that is present even at $0\%$ pruning which reduces noise for the federated aggregation. 

Each architecture has a unique multi-dimensional tensor as shown in Table~\ref{tbl:compare-archs}. 
These results empirically support the selection of multi-dimensional DCT-$IV$ for \fspace transformation. 

\subsection{Effect of \feddct on Computation Overheads}
\label{sec:cost-compare}

Understanding the computation overheads is essential for FL methods, particularly in environments with limited computing resources. To assess the impact of \feddct, we compared it against baseline methods over 100 communication rounds across all the datasets.
The results showed that \feddct requires up to a 6\% increase in resources compared to \fedsim and \fedavg.
However, this overhead is less than 5\% when compared to \fedprox. In our setup with a 1.7 GHz Quad-Core CPU, a 6\% increase amounted to an additional 0.03 seconds of computation time.
This increase is relatively insubstantial when weighed against the benefits that \feddct offers.
Therefore, the slight increase in computation can be neglected when weighed against the enhanced communication efficiency it provides, saving network resources and overall efficiency.

\subsection{Analysing the Effect of Non-IID on \feddct}
\label{sec:noniid-compare}

\begin{figure}[htb]
\centering
\includegraphics[width=0.85\textwidth]{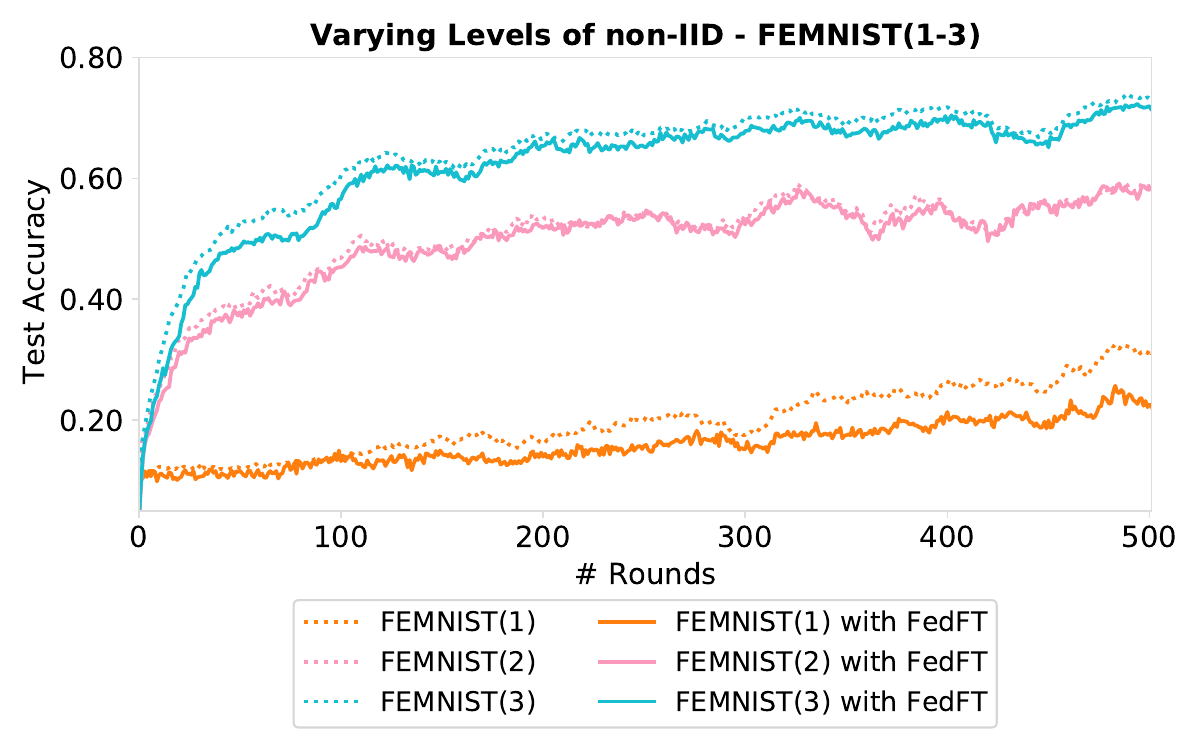}
\caption{Varying levels of non-IID with three versions of the FEMNIST dataset}
\label{fig:varying-non-iid}
\end{figure}

FL environments inherently support and often require the handling of non-IID data due to their distributed nature. This experiment is focused on evaluating how \feddct performs under different levels of non-IID data.
Figure \ref{fig:varying-non-iid} illustrates the outcomes obtained from the three FEMNIST variants, representing varying levels of non-IID, when applied to the \fedavg baseline. 
In the figure, two types of lines are used to represent the results. The solid lines show how \fedavg, combined with \feddct, performs. In contrast, the dashed lines show the performance of the standard \fedavg method. 
The presented plots depict the average results obtained from 35 independent runs with random seeds conducted over 500 communication rounds.
Our observations indicate that, in the experiment, \feddct consistently maintains comparable performance across all levels of non-IID. Additionally, we note that any initial decrease in performance seen in FEMINST(2) and FEMNIST(3) gradually recovers in the later rounds. Additionally, it is essential to highlight that the overall performance of \fedavg in the FEMINST(1) dataset is comparatively weaker, requiring more communication rounds for convergence compared to the baseline \fedavg. 

However, we observe that \feddct can catch up and follow a similar convergence trend in this extreme non-IID scenario.
\feddct still shows a consistent trend, even in these varied non-IID conditions. 
This detailed investigation and generalisability studies confidently suggest that \feddct is appropriately suited for non-IID settings in Federated Learning.

\subsection{Impact of \feddct Pruning}
\label{sec:compare-prune}

\begin{figure*}[htb]
\centering
\includegraphics[width=\textwidth]{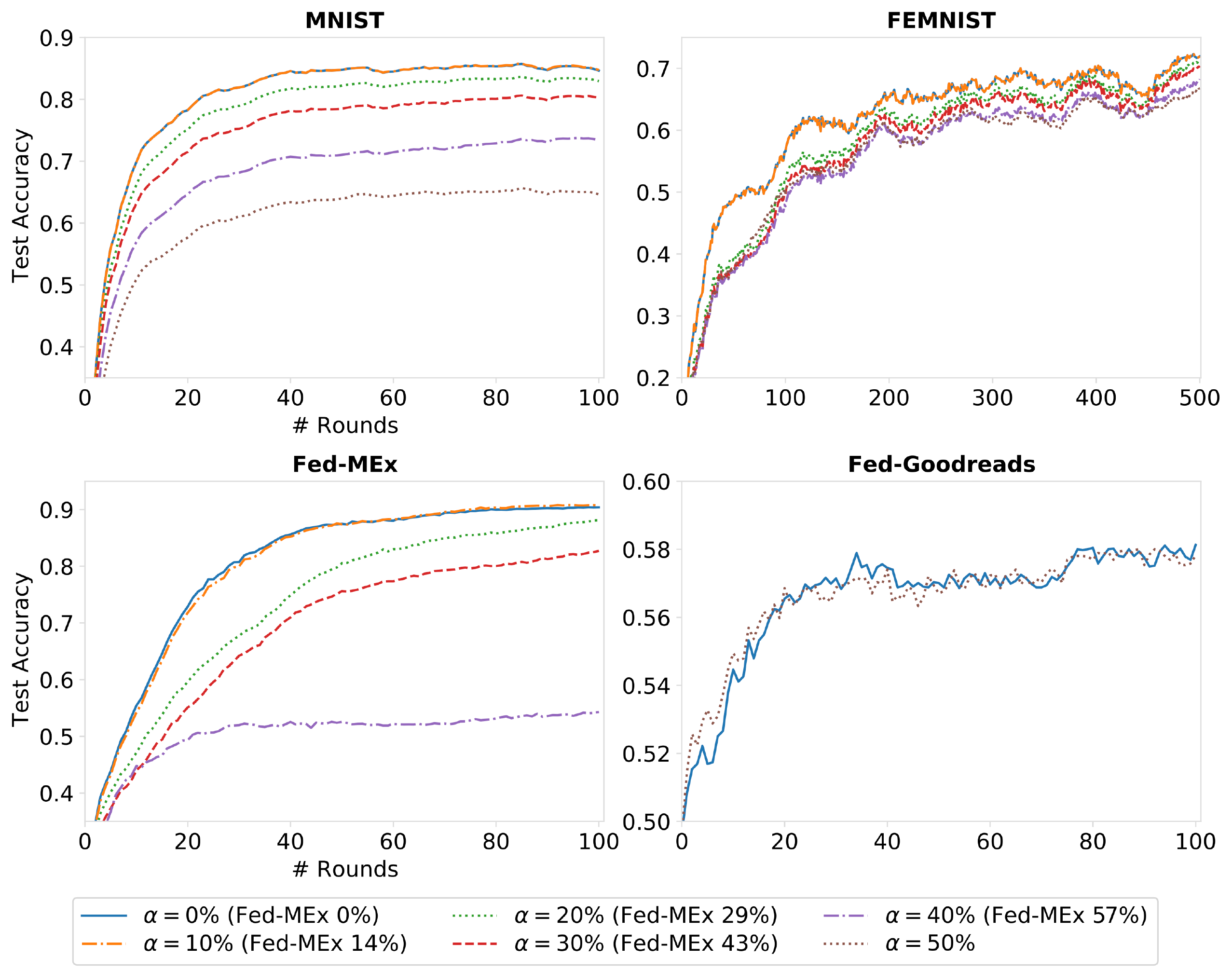}
\caption{\feddct with pruning }
\label{fig:compare-prune-0}
\end{figure*}

The ability to compress model parameters using pruning or quantisation~(such as with JPEG images and video streaming) is a crucial aspect of communication in \fspace. 
We examined the extent to which pruning can compress while preserving performance. 
Figure~\ref{fig:compare-prune-0} presents test accuracy with increasing values of the pruning rate~$\alpha$ for each dataset. 
As expected, accuracy suffers with higher values of $\alpha$.
This poor performance is mostly evident for pruning with $\alpha > 20\%$. 
Note that the model's inability to overcome the negative impact of high pruning on its performance prior to convergence results in a sub-optimal test accuracy post-convergence.

However, it is encouraging to observe that at lower levels of pruning, comparable performance to no-pruning is achieved. 
This suggests that there is a sweet-spot where pruning can achieve comparable or in some cases better accuracy than no-pruning. 
For instance, test accuracy with $\alpha=10\%, 10\%$ and $14\%$ is comparable to $\alpha=0\%$ with MNIST, FEMNIST and \mex datasets respectively. 
The most favourable outcomes with pruning are observed in \greads, where $\alpha=50\%$ yields performance comparable to that of no-pruning across communication rounds. This finding suggests that the magnitudes of high-frequency coefficients (i.e., those preserved without pruning) unintentionally carried noisy information, which initially hindered the federated aggregation.

\subsection{Communication Efficiency with \feddct}

\begin{figure*}[htb]
\centering
\includegraphics[width=\textwidth]{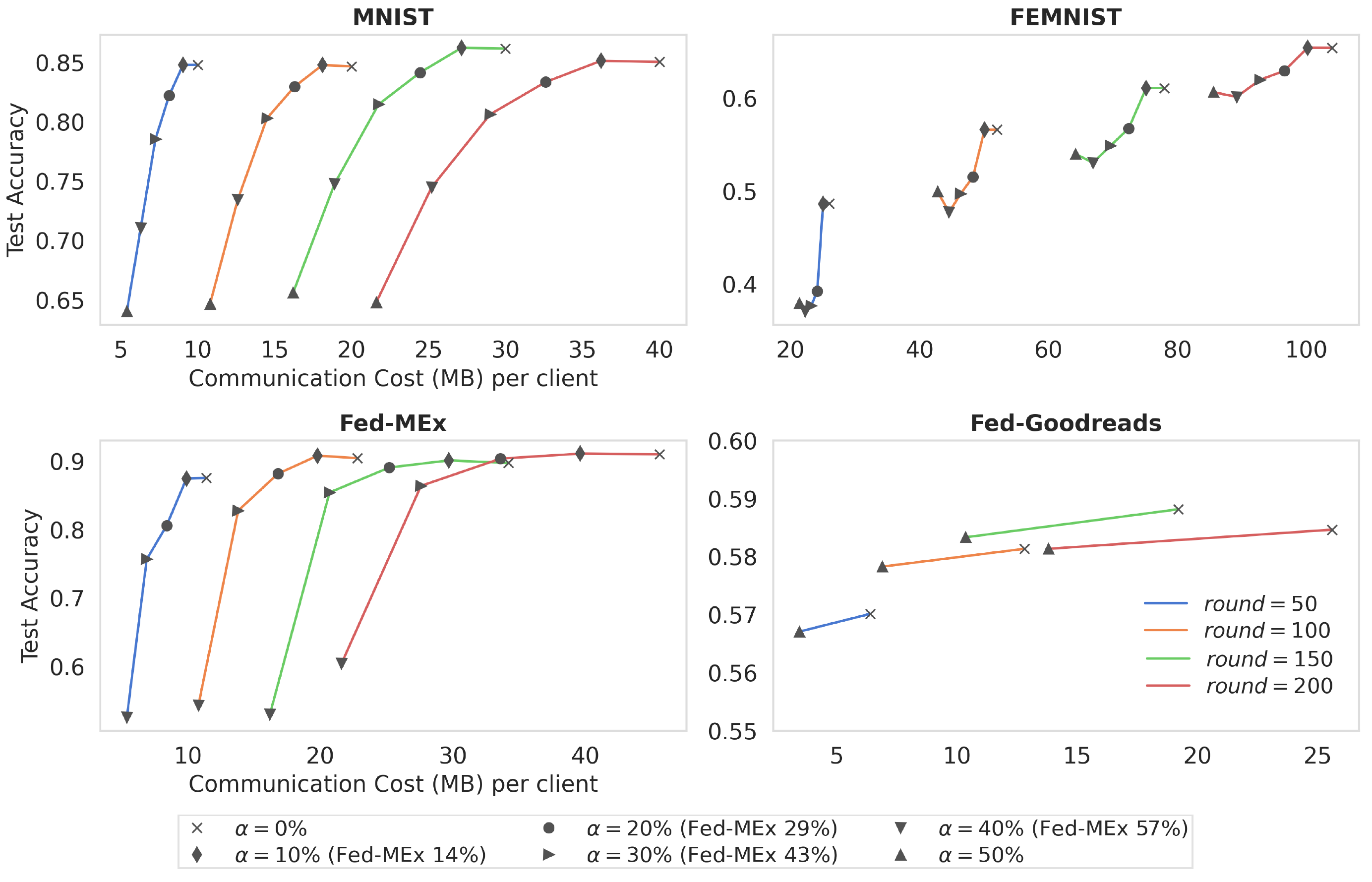}
\caption{Optimising the upstream communication cost with \feddct}
\label{fig:compare-communication-cost}
\end{figure*}

To study the communication efficiency, we plot upstream communication costs in Figure~\ref{fig:compare-communication-cost}. 
Here, a single trend line of a plot shows the test accuracy values measured at a particular communication round~(coloured lines). 
The x-axis is the accumulated upstream communication cost per client in MB measured on different $\alpha$ indicated by the markers.

\begin{table*}[htb]
\caption{Communication costs and accuracy for each pruning $\alpha$ Percentage at the 200th round}
\centering
\small
\renewcommand{\arraystretch}{1.1}
\begin{tabular}{p{8mm}|p{10mm}p{10mm}|p{10mm}p{10mm}|p{10mm}p{10mm}|p{10mm}p{10mm}}
% \toprule
\multicolumn{1}{l}{} &
  \multicolumn{2}{c}{\begin{tabular}[c]{@{}c@{}}\textbf{MNIST}\end{tabular}} &
  \multicolumn{2}{c}{\begin{tabular}[c]{@{}c@{}}\textbf{FEMNIST}\end{tabular}} &
  \multicolumn{2}{c}{\begin{tabular}[c]{@{}c@{}}\textbf{\mex}\end{tabular}} &
  \multicolumn{2}{c}{\begin{tabular}[c]{@{}c@{}}\textbf{Fed-Goodreads}\end{tabular}} \\ \hline

$\alpha$ & Cost (MB)  &  Acc.      & Cost (MB) & Acc.    & Cost (MB)    & Acc. & Cost (MB)   & Acc.    \\
\hline
0\%   & 40          & 85\%          & 104          & 65\%          & 45.6          & 90\%          & 25.6          & 58\%          \\ \hline

10\%  & 36.2          & 85\%          & \textbf{100.2} & \textbf{65\%} &              &               &              & \textbf{}     \\ \hline

14\%  &              &               &               &               & 39.6          & 91\%          &              &               \\ \hline

20\%  & \textbf{32.6} & \textbf{83\%} & 	96.6          & 62\%          & \textbf{}    & \textbf{}     &              &               \\ \hline

29\%  &              &               &               &               & \textbf{33.6} & \textbf{90\%} &              &               \\ \hline

30\%  & 29          & 80\%          & 92.8          & 61\%          &              &               &              &               \\ \hline

40\%  & 25.2          & 74\%          & 89.2          & 60\%          &              &               &              &               \\ \hline

43\%  &              &               &               &               & 27.6          & 86\%          &              &               \\ \hline

50\%  & 21.6          & 64\%          & 85.6          & 60\%          &              &               & \textbf{13.8} & \textbf{58\%} \\ \hline

57\%  &           &           &           &          &      21.6         &            60\%    & & 
\\ \hline
\end{tabular}
\label{tbl:upstreamcost} 
\end{table*}

Firstly, Figure~\ref{fig:compare-communication-cost} confirms the general finding in Figure~\ref{fig:compare-prune-0} that accuracy with pruning in the range, $0 < \alpha < 20\%$, is comparable to no pruning ($\alpha = 0$).
Secondly, we can observe how \feddct pruning can optimise communication cost when given thresholds for test accuracy and communication rounds. 
The values shown in Figure \ref{fig:compare-communication-cost} are outlined in Table~\ref{tbl:upstreamcost} at the 200th round (i.e. red color line). The \textbf{bolded} figures highlight the optimal balance between accuracy and communication efficiency.

Finally, we address the issue where pruning at early stages of model training can lead to sub-optimal test accuracy. 
To mitigate the risk of losing information about clients at the early stages of training, we propose applying pruning after some communication rounds, preferably post-convergence. Post-convergence pruning can enhance communication efficiency by allowing the fine-tuning of a model after convergence. We choose these two datasets (MNIST, \mex) due to their apparent convergence, enabling us to establish the pruning threshold. When applying pruning, MNIST performances across all $\alpha$ values are comparable to no pruning ($\alpha=0\%$). 
\mex also maintains comparable performances up to $\alpha=43\%$. 
We attribute these improved pruning performances to the reduced magnitudes of weight adjustments made by client models after the global model converges.

\subsection{Impact of \feddct Pruning on \fedsim}

\begin{figure*}[htb]
\centering
\includegraphics[width=\textwidth]{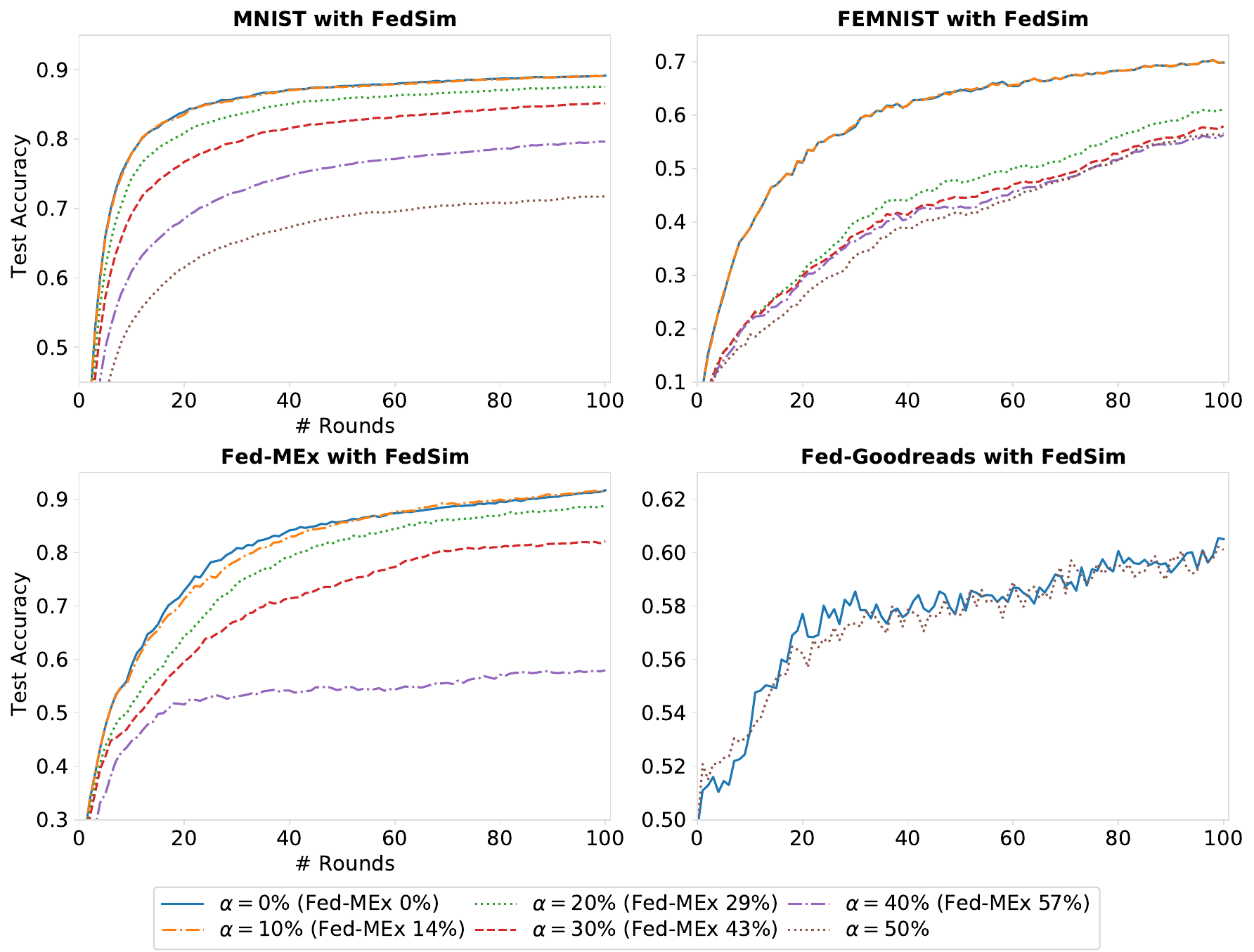}
\caption{\feddct with pruning on \fedsim}
\label{fig:compare-prune-fedsim}
\end{figure*}

Building upon the analysis in Figure~\ref{fig:compare-prune-0}, we further explore the balance between pruning and performance retention, specifically within the \fedsim \citep{fedsim} aggregation methodology.
Figure~\ref{fig:compare-prune-fedsim} presents test accuracy with increasing values of the pruning rate~$\alpha$ for each dataset with \feddct applied on \fedsim. 
We note that at a pruning rate of $\alpha = 10\%$ (\mex: $\alpha = 14\%$), \feddct achieves comparable performance, enhancing communication efficiency.

As expected, the accuracy declines with higher values of $\alpha$.
However, it is significant to observe that, despite this reduction in accuracy, the core performance benefits of the \fedsim method remain largely intact.
This resilience highlights the robustness of the \feddct pruning approach, particularly in synergy with \fedsim advanced aggregation strategy.
We specifically chose to test \feddct with the \fedsim method to explore its adaptability and performance in personalised/clustered FL algorithms. 
This approach is particularly relevant for real-world applications, where similarities among clients play a crucial role in enhancing the efficiency and effectiveness of the learning process.

\subsection{Impact of \feddct Pruning Post-convergence}

\begin{figure}[htb]
\centering
\includegraphics[width=0.8\columnwidth]{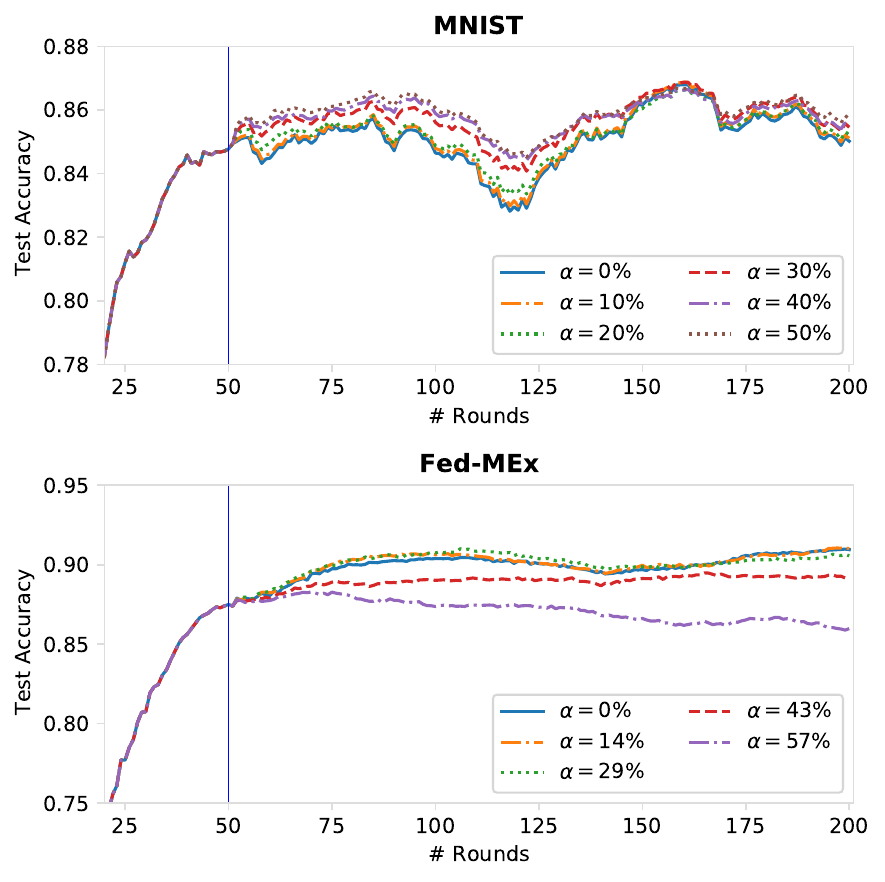}
\caption{\feddct post-convergence pruning (round$>50$)}
\label{fig:compare-prune-post}
\end{figure}

In FL environments, learning often occurs in incremental steps involving a substantial number of clients and rounds of communication. This process can continue to improve model performance even after initial convergence. We study post-convergence pruning in  Figure~\ref{fig:compare-prune-post}, where we plot the results with a pruning threshold set at 50 communication rounds (represented by the blue vertical line) for MNIST and \mex. We chose these two datasets due to their apparent convergence, which enabled us to establish the pruning threshold.

When applying pruning, MNIST performances across all $\alpha$ values are comparable to no pruning  ($\alpha=0\%$). \mex also maintains comparable performances up to $\alpha=43\%$. We attribute these improved pruning performances to the reduced magnitudes of weight adjustments made by client models after the convergence of the global model.
These findings suggest that post-convergence pruning can effectively maintain model performance while optimising communication efficiency in FL settings. 

\section{Conclusion}
\label{sec:conc}
\feddct introduced a novel FL methodology that communicates model parameters in \fspace and performs federated aggregation in that same space.
DCT-$IV$ transformed and pruned model parameters of \feddct achieved reduced communication costs while maintaining model accuracy.
Extensive experiments conducted on four FL datasets 
and employing three state-of-the-art FL methodologies demonstrate the generalisability of \feddct across diverse neural model architectures and FL methodologies.
\feddct is a generalisable solution, achieving communication savings of $5\%-30\%$ while maintaining comparable accuracy.
A promising direction for future research is to dynamically identify the 'sweet spot' for optimal pruning results by analysing training patterns. 
Additionally, a limitation to be explored in the future is the impact of varying levels of non-IID data on \feddct and the associated security implications of implementing \feddct in FL systems.

\bibliographystyle{elsarticle-num}
\bibliography{references}

\clearpage
\appendix
\begin{center}
\textbf{{\Large Appendix}}
\end{center}
\section{\feddct adaptations of FL methodologies}

\label{ap:algos}
\begin{algorithm}[htb!]
\caption{\feddct adaptation of \fedsim}
\label{algo:ftfedsim}
\begin{algorithmic}[1]
\REQUIRE $w_0$: initial global model, \textcolor{blue}{$\alpha$: Pruning Rate}, $K$: num. of selected clients
\REQUIRE \textcolor{blue}{$T(.)$ DCT Function, $\hat{T}(.) $ Inverse DCT Function, $P(.)$ Pruning Function}
\textcolor{blue}{\STATE ${\hat{w}_0} =  T(w_0) \gets$ DCT transformation}
\FOR{t=1,2,..}
\STATE Broadcast $\hat{w}_{t}$ to all clients
\STATE Select $\mathcal{S}$ clients where $\mathcal{S} \subset \mathcal{K}$
% \STATE $G$ $\leftarrow$ $g_1$...$g_{|\mathcal{S}|}$, where gradients $g_k$ received from each client k $\in$ $S$
\STATE $\mathcal{C}$ $\leftarrow$ Clustering($S$, \textit{n\_clusters})
\FORALL{$c \in \mathcal{C}$ }
    \FORALL{$k \in c$}  
        \STATE  \textcolor{blue}{$w^k_t = \hat{T}(\hat{w}_t) \gets$ inverse DCT transform}
        \STATE $w^k_{t+1} \gets$ updates $w^k_t$ using SGD 
        \STATE \textcolor{blue}{$\Delta w_{t+1}^k = w_{t+1}^k - w^k_t \gets $ \textit{ update differences}}
        \STATE \textcolor{blue}{$\Delta \hat{w}_{t+1}^k = P(T(\Delta w_{t+1}^k), \alpha) \gets $DCT transform and prune}
        \STATE \textcolor{blue}{Send $\Delta \hat{w}_{t+1}^k$ to the server}
    \ENDFOR
\STATE \textcolor{blue}{$\hat{\bar{w}}^c_{t+1} \gets$ ClusterAggregation($\{\hat{w}_{t}+\Delta \hat{w}^k_{t+1}\}\text{ } \forall\text{ }k\text{ }\in\text{ }c$)}
\ENDFOR
% \STATE $\hat{w}_{t+1} \gets$ GlobalAggregation($\hat{\bar{w}}^1_{t+1}, \hat{\bar{w}}^{2}_{t+1},..., \hat{\bar{w}}^{|\mathcal{C}|}_{t+1}$)
\STATE $\hat{w}_{t+1} \gets$ GlobalAggregation($\{\hat{\bar{w}}^c_{t+1}\}\text{ }\forall\text{ }c\text{ }\in\text{ }\mathcal{C}$)
\ENDFOR
\end{algorithmic}
\end{algorithm}

\begin{algorithm}[htb!]
\caption{\feddct adaptation of \fedprox}
\label{algo:ftfedprox}
\begin{algorithmic}[1]
\REQUIRE $w_0$: initial global model, \textcolor{blue}{$\alpha$: Pruning Rate}, $K$: num. of selected clients
\REQUIRE \textcolor{blue}{$T(.)$ DCT Function, $\hat{T}(.) $ Inverse DCT Function, $P(.)$ Pruning Function}
\textcolor{blue}{\STATE ${\hat{w}_0} =  T(w_0) \gets$ DCT transformation}
\FOR{t=0,1,2, ...}
\STATE Broadcast $\hat{w}_{t}$ to all clients
\STATE Select $K$ clients with probability $p_k$
    \FORALL{$k \in K$}  
        \STATE  \textcolor{blue}{$w_t^k = \hat{T}(\hat{w}_t) \gets$ inverse DCT transform}
        \STATE  $w^k_{t+1} \gets$ update $w^k_t$ using $F_k(w) + \frac{\mu}{2}\Vert{w-w^t}\Vert^2$ (\cite{fedprox})
        \STATE \textcolor{blue}{$\Delta w_{t+1}^k = w_{t+1}^k - w^k_t \gets $ \textit{ update differences}}
        \STATE \textcolor{blue}{$\Delta \hat{w}_{t+1}^k = P(T(\Delta w_{t+1}^k), \alpha) \gets $ DCT transform and prune}
        \STATE \textcolor{blue}{Send $\Delta \hat{w}_{t+1}^k$ to the server}
    \ENDFOR
    \STATE \textcolor{blue}{$\hat{w}_{t+1} \gets \sum_{k=1}^{K} \frac{n_k}{n}(\hat{w}_{t}+ \Delta \hat{w}_{t+1}^k )$ \textit{Federated Aggregation on update differences}}
    \STATE \textcolor{blue}{Set PGD Parameters $\gets \hat{T}(\hat{w}_{t+1})$  }
\ENDFOR
\end{algorithmic}
\end{algorithm}

\end{document}